\documentclass[journal]{IEEEtran}
\usepackage{graphicx}

\usepackage{color}
\usepackage{changepage}
\usepackage{times}
\usepackage{float}
\usepackage{amssymb}
\usepackage{multirow}
\usepackage{bbding}

\usepackage{subfigure}
\usepackage{booktabs}
\usepackage{subfigure}
\usepackage{multirow}
\usepackage{amsmath,epsfig}
\usepackage{cite}
\usepackage{url}
\usepackage{color}

\usepackage{algorithm}
\usepackage{algorithmic}
\usepackage{rotating}
\ifCLASSINFOpdf
\else
\fi
\begin{document}
%
\title{Blind Quality Assessment for in-the-Wild Images via Hierarchical Feature Fusion and Iterative Mixed Database Training}
%
%
%

\author{Wei~Sun,
	Xiongkuo~Min,~\IEEEmembership{Member,~IEEE,}
	Danyang~Tu,
	Siwei~Ma,~\IEEEmembership{Senior Member,~IEEE,}
	and Guangtao~Zhai,~ \IEEEmembership{Senior Member,~IEEE}
\thanks{Manuscript received August 25, 2022; revised March 23, 2023; accepted April 09, 2023. This work was supported in part by Shanghai Municipal Science and Technology Major Project 2021SHZDZX0102, in part by the National Natural Science Foundation of China under Grants 62225112, 61831015, 62271312, in part by the Fundamental Research Funds for the Central Universities, in part by the Shanghai Pujiang Program under Grant 22PJ1407400, and in part by National Key R\&D Program of China 2021YFE0206700. This work was presented in part at the 2022 IEEE International Symposium on Broadband Multimedia Systems and Broadcasting \cite{sun2022blind}. The guest editor coordinating the review of this article and approving it for publication was Dr. Aladine Chetouani. (\emph{Corresponding author: Xiongkuo Min and Guangtao Zhai.})}
\thanks{W. Sun, X. Min, D. Tu, and G. Zhai are with the Institute of Image Communication and Information Processing, Shanghai Jiao Tong University and G. Zhai are also with the MoE Key Lab of Artificial Intelligence, AI Institute, Shanghai Jiao Tong University, Shanghai 200240, China (email:\{sunguwei, minxiongkuo, danyangtu, zhaiguangtao\}@sjtu.edu.cn)}
\thanks{S. Ma is with the Institute of Digital Media, School of Electronic Engineering and Computer Science, Peking University, Beijing 100871, China (e-mail: swma@pku.edu.cn).}}
\maketitle

\begin{abstract}
Image quality assessment (IQA) is very important for both end-users and service providers since a high-quality image can significantly improve the user's quality of experience (QoE) and also benefit lots of computer vision algorithms. Most existing blind image quality assessment (BIQA) models were developed for synthetically distorted images, however, they perform poorly on in-the-wild images, which are widely existed in various practical applications. 
In this paper, we propose a novel BIQA model for in-the-wild images by addressing two critical problems in this field: \textit{how to learn better quality-aware feature representation}, and \textit{how to solve the problem of insufficient training samples in terms of their content and distortion diversity}. Considering that perceptual visual quality is affected by both low-level visual features (e.g. distortions) and high-level semantic information (e.g. content), we first propose a staircase structure to hierarchically integrate the features from intermediate layers into the final feature representation, which enables the model to make full use of visual information from low-level to high-level. Then an iterative mixed database training (IMDT) strategy is proposed to train the BIQA model on multiple databases simultaneously, so the model can benefit from the increase in both training samples and image content and distortion diversity and can learn a more general feature representation. Experimental results show that the proposed model outperforms other state-of-the-art BIQA models on six in-the-wild IQA databases by a large margin. Moreover, the proposed model shows an excellent performance in the cross-database evaluation experiments, which further demonstrates that the learned feature representation is robust to images with diverse distortions and content. The code is available at \url{https://github.com/sunwei925/StairIQA}.
\end{abstract}


\IEEEpeerreviewmaketitle

\section{Introduction}

\IEEEPARstart{W}{ith} the advent of the mobile era, billions of images are generated in various social media applications every day, most of which are captured by amateur users in various in-the-wild environments. Different from pictures shot by photographers, the quality of ordinary-user-generated images is often degraded by distortions like under/over exposure, low visibility, motion blur, ghosting, etc. A high-quality image on one hand can improve the viewer's Quality of Experience (QoE) and on the other hand can benefit lots of computer vision algorithms. Therefore, the image quality is a very important indicator for the service providers to deliver high-quality content to users and for the visible light camera based systems to filter low-quality images to avoid decision errors. With massive numbers of images being generated by millions of cameras every moment, there is an urgent need to develop an effective quality assessment model for in-the-wild images.

According to the amount of reference image information needed, image quality assessment (IQA) algorithms can be classified into full-reference IQA (FR IQA), reduced-reference IQA (RR IQA), and no-reference IQA (NR IQA), also widely known as blind IQA (BIQA). These IQA methods have been widely applied in many fields such as image compression \cite{zhang2020data}, image restoration \cite{wu2020subjective,zhai2021perceptual}, virtual reality \cite{zhou2021omnidirectional}, etc. Due to the lack of pristine images, only BIQA models are qualified for evaluating the quality of in-the-wild images. Previous BIQA models \cite{moorthy2011blind, mittal2012no,gu2014using,min2017blind,min2018blind,zhai2019free} mainly focus on images with synthetic distortions such as JPEG compression, Gaussian blur, etc. However, it should be noted that the difference between images with synthetic and authentic distortions is quite large, which is listed as follows:
 
First, synthetic distortions, including their types and distortion degrees, are more regular than authentic distortions. For example, previous studies \cite{moorthy2011blind,saad2012blind,mittal2012no} find that natural scene statistics (NSS) can effectively describe different kinds of synthetic distortions. It is not surprising since synthetic distortions are generated by specific algorithms, while authentic distortions are produced during the shooting process, which may be affected by many factors such as shooting environment, equipment, photographing techniques, and usually contain mixtures of multiple uneven distortions.
Second, synthetic distortions are usually global uniform (e.g. JPEG compression in Fig. \ref{example_images} (a) and white noise in Fig. \ref{example_images} (b)) because these distortions are added to the whole images, however, authentic distortions not only can be global uniform (e.g. low illumination, Fig. \ref{example_images} (c)), but also can be non-uniform (e.g. object moving, Fig. \ref{example_images} (d)).
Third, the synthetically distorted image is degraded from a perfect image, whose content is also often of high quality, so the image content is not the primary factor that decides the image quality when compared with their distortion type and degree. But for authentically distorted images, we noticed that the image quality is closely related to its content (i.e. captured objects, image composition, etc.), which is also called aesthetic characteristics of the image \cite{wu2022disentangling}. As a result, though existing BIQA models \cite{moorthy2011blind, mittal2012no,gu2014using,min2017blind,min2018blind,zhai2019free} have achieved remarkable performance on synthetically distorted images, there is still a great challenge to assess the quality of in-the-wild images.
\begin{figure}[!t]
	\centering
	\includegraphics[height=2.5in]{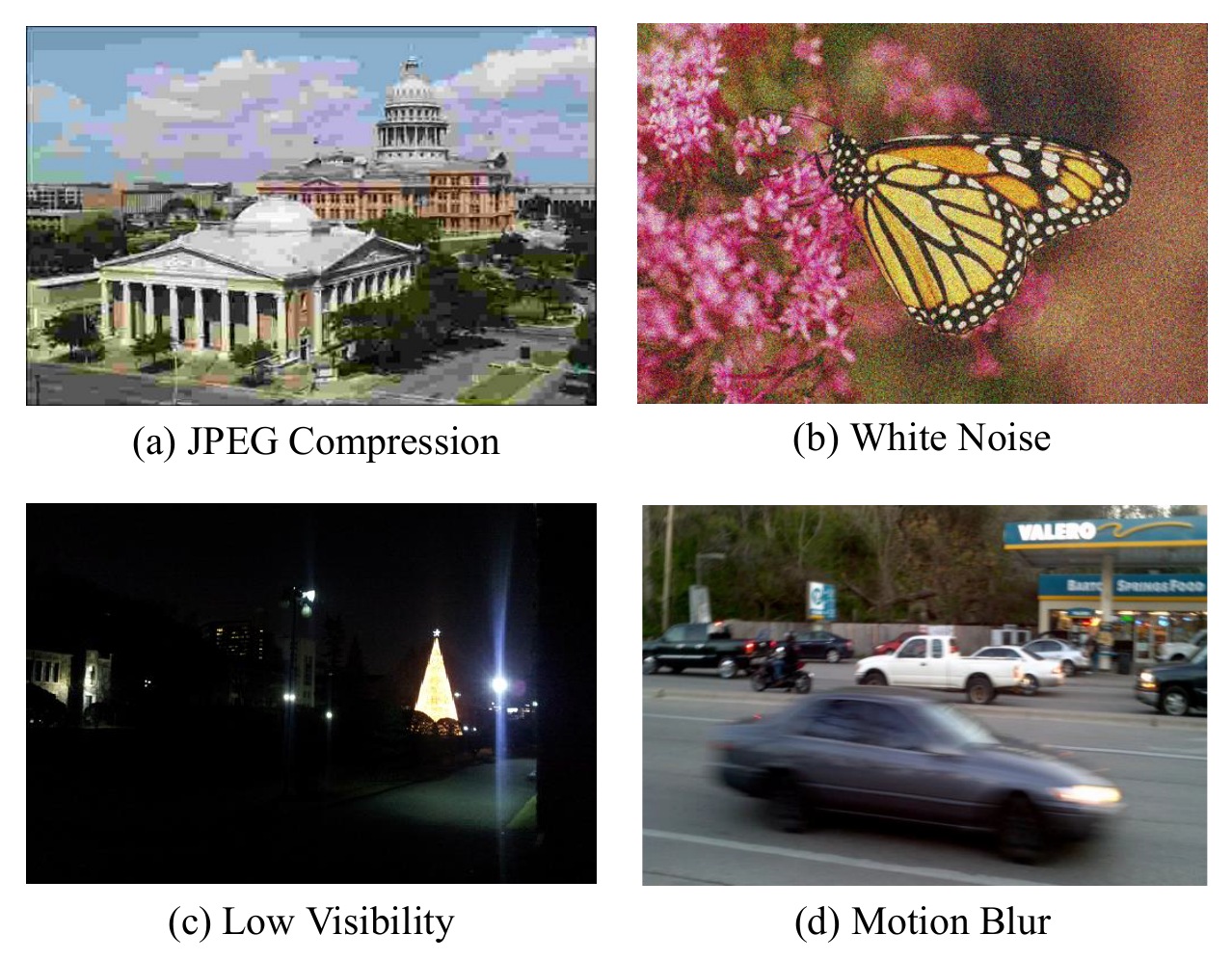}
	\caption{Examples of images with synthetic distortions and authentic distortions. (a) and (b) are synthetically distorted images, (c) and (d) are authentically distorted images.}
	\label{example_images}
\end{figure}

Existing BIQA models generally follow such routine: 1) extracting quality-aware features, and 2) mapping these features into the quality score via a regression model. Commonly used quality-aware features include NSS features \cite{moorthy2011blind, mittal2012no}, free energy features\cite{gu2014using,zhai2019free}, corners/textures\cite{min2018blind}, etc., while commonly used regression models include support vector regression, random forest regression, etc. Recently, deep learning technologies show great ability to solve various visual signal processing problems. The latest BIQA models adopt the deep learning based architecture, which utilizes a deep convolutional neural network (CNN) to extract quality-aware features of distorted images, and then regresses them to quality scores with a fully connected network. This kind of architecture allows it to be trained in an end-to-end manner and has been dominant in the BIQA fields. 

Compared with handcrafted features, features extracted by CNN are more powerful and more suitable for in-the-wild images. However, there exist two challenges for training a deep CNN model for BIQA. First, the commonly used backbone networks such as VGG \cite{simonyan2015very}, Resnet \cite{he2016deep}, etc. are designed for the image classification task, where the extracted features are at the semantic level, however, the perceptual visual quality is affected by both high- and low-level visual features. So, it is not an optimal option to directly use a popular CNN architecture as the backbone of BIQA tasks due to the loss of low-level feature information.
Second, deep CNN models contain millions of parameters and require a large scale of diverse samples for training. Since subjective IQA experiments are extremely time-consuming and cumbersome, the scale of existing IQA databases is relatively small and the diversity of image content and distortion is also insufficient in these databases, making it difficult to train an excellent deep model. The patch-based training method \cite{kang2014convolutional} is a naive solution to augment the database, which divides an image into many patches and the quality score of each patch is set to be the same as the corresponding image. This strategy is useful for synthetic distortions, but does not work for authentic distortions since the distortions of authentically distorted images are not global uniform. What's more, the patch-based method cannot increase the diversity of image content. A more detailed discussion about the relationship between the quality of the whole image and its patches can be referred to \cite{ying2020from}.

In this paper, we propose a novel BIQA model for in-the-wild images, which tries to solve the above two challenges. First, we propose a staircase structure to hierarchically incorporate the features from intermediate layers into the final feature representation, which makes the model learn more effective quality-aware feature representation. Previous studies \cite{zeiler2014visualizing, ranjan2017hyperface} indicate that the features extracted from different stages of a CNN model represent different visual information. For example, the features extracted from bottom convolution layers correspond to low-level information such as edges and corners, while the features extracted from top convolution layers are at the semantic level. Through fusing the features from intermediate layers, the CNN model can fully utilize the visual information from low-level to high-level and learn better feature representations for quality evaluation.

Second, considering that existing IQA databases are relatively small for training a deep CNN model, we propose an iterative mixed database training strategy (IMDT) to train the BIQA model on multiple databases simultaneously, which can take full advantage of diverse image content and distortions of multiple databases and make the model learn a more general feature representation. As stated above, a BIQA model can be divided into feature extraction and quality regression modules. For different IQA databases, the absolute quality scores are not directly comparable since their experiment setups, participants, and evaluation criteria are not the same, but their relative quality scores are consistent in each database. The motivation of the IMDT strategy is that quality-aware features extracted from BIQA models should be universal for all IQA databases, and the different quality scales of each database can be aligned by the quality regression modules. 
The IMDT strategy allows the feature extraction module to be trained on all databases and the quality regression module to be trained by the corresponding target databases. Hence, the feature extraction module can benefit from the increase in training samples and image content and distortion diversity, which is particularly important for small IQA databases.

In summary, this paper has made the following contributions.
\begin{enumerate}
	\item We propose a staircase structure to hierarchically integrate the features from intermediate layers into the final feature representation. The staircase structure can be integrated with any popular CNN backbones, and make the model learn better feature representations for BIQA tasks. 
	\item We propose an iterative mixed database training strategy (IMDT) to train the BIQA model on multiple databases, which on one hand can alleviate the problem of insufficient training samples, and on the other hand can learn more general quality-aware features from the diverse image content and distortions of multiple databases.
	\item Experimental results show that the proposed model achieves the best performance on six in-the-wild IQA databases, and also achieves an excellent performance in the cross-database evaluation, which demonstrate the effectiveness and generalizability of the proposed model.
\end{enumerate}


\begin{figure*}[!t]
	\centering
	\includegraphics[height=3in]{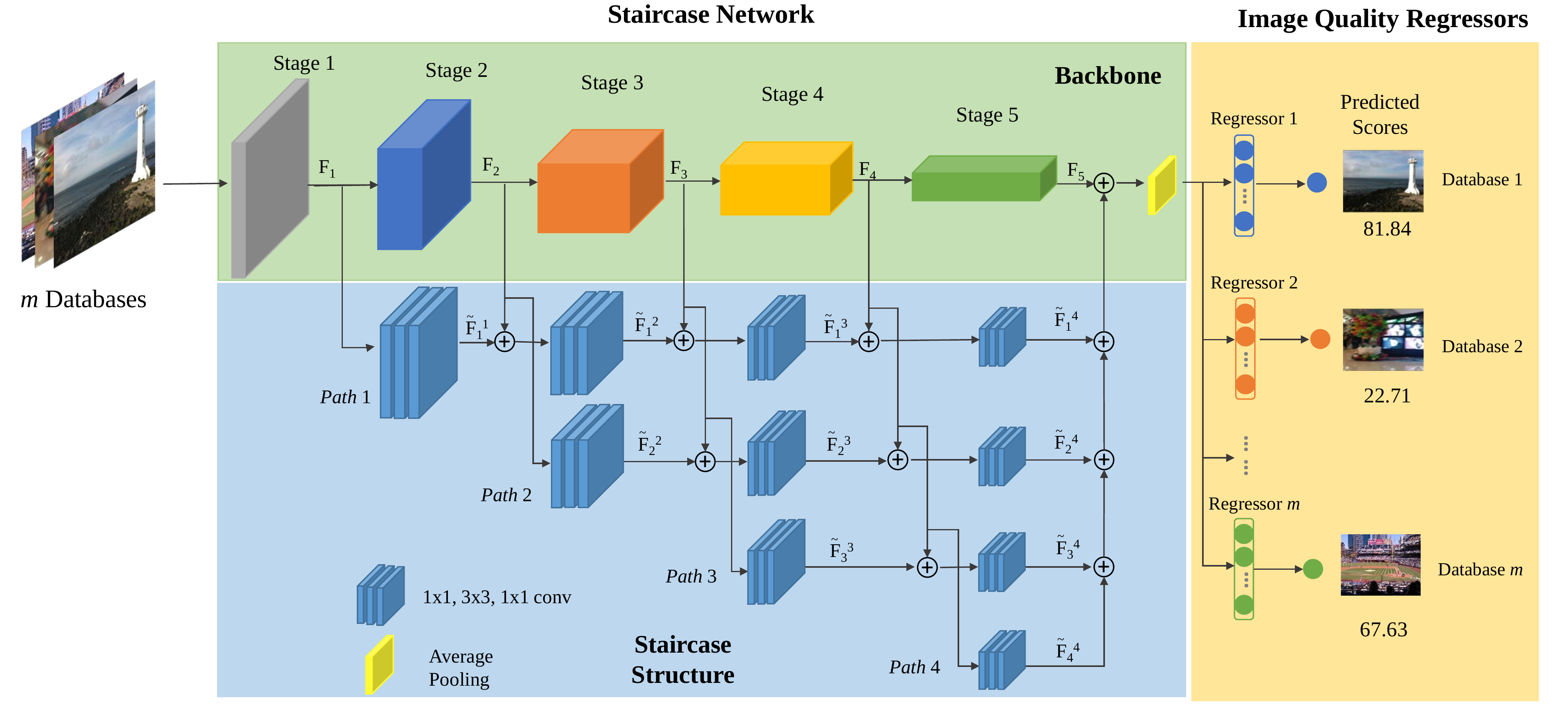}
	\caption{The network architecture of the proposed model. The proposed model includes the staircase network for quality-aware feature extraction and image quality regressors for mapping the quality-aware features to quality score spaces. $N$ image quality regressors correspond to $N$ IQA databases that need to be trained simultaneously. Here, we assume that there are four stages in the backbone network.}
	\label{framework}
\end{figure*}

\section{Related Work}
\label{related_work}
In this section, we briefly review the research on BIQA and mixed database training.

\subsection{Blind Image Quality Assessment}
\label{biqa}
Based on the methods of feature extraction, BIQA models can be divided into two categories: handcrafted feature based and learning feature based. Natural scene statistic (NSS) is a common handcrafted feature for BIQA models. The motivation of NSS-based models is that high-quality natural scene images obey certain statistical properties, while quality degradations cause images to deviate from these statistics. For example, DIIVINE \cite{moorthy2011blind} first identifies the distortion type of the image, and then conducts distortion-specific IQA using NSS features extracted in the wavelet domain. BLIINDS-II \cite{saad2012blind} utilizes NSS of discrete cosine transform (DCT) coefficients. BRISQUE \cite{mittal2012no} uses the scene statistics of local luminance coefficients to quantify possible losses of ``naturalness''. 
Besides NSS, other handcrafted features such as free energy \cite{zhai2019free}, image texture, corner/edge, etc. are also used in previous BIQA models. Gu \textit{et al.} \cite{gu2014using} develop a NR free energy based robust metric (NFERM) using three groups of features: features extracted by the free energy model, image structure and gradient features, and NSS features of the mean subtracted contrast normalized coefficients. Min \textit{et al.} \cite{min2018blind}  integrate the similarities of corners and local binary patterns between distorted images and the corresponding pseudo images as the quality score, where the pseudo images can be in multiple distortion levels. Liu \textit{et al.} \cite{liu2019unsupervised} propose an unsupervised BIQA model by quantifying the distortions from the structure, naturalness, and the perception quality variations from the pristine natural images.

Learning feature based BIQA methods \cite{kang2014convolutional,xu2016blind,xu2016multi,bosse2017deep,ma2017end,zhang2020blind,su2020blindly,zhu2021generalizable,ke2021musiq} utilize the machine learning technologies to automatically learn features from IQA databases. Early studies usually use codebook learning methods. For example, Ye \textit{et al.} \cite{ye2012unsupervised} utilize soft-assignment coding with max pooling to derive image representations for quality assessment. Zhang \textit{et al.} \cite{zhang2014training} extract log-Gabor responses to formulate codebook by sparse filer learning. Xu \textit{et al.} \cite{xu2016blind} use high order statistics information to reduce the number of the codebook and improve the performance. 

Since deep neural networks show great learning ability in recent years, it has become a trend to use deep CNN to blindly evaluate the image quality.
Kang \textit{et al.} \cite{kang2014convolutional} use a shallow CNN model consisting of one convolutional layer and two fully connected layers to estimate the quality of small patches, and then the image level quality score is averaged by the evaluated patch scores of the corresponding image.
Bosse \textit{et al.} \cite{bosse2017deep} further deepen the CNN model by jointly learning the quality and weight of patches, where the weight is the relative importance of the patch quality to the global quality estimation.
Ma \textit{et al.} \cite{ma2017end} propose a multi-task deep neural network for BIQA by joint identifying distortion types and assessing the image quality.
Zhang \textit{et al.} \cite{zhang2020blind} propose to merge features extracted from two kinds of CNN models into a better representation by bilinearly pooling, where two CNN are respectively pretrained on the distortion type and level classification task and the image classification task.
Su \textit{et al.} \cite{su2020blindly} develop a self-adaptive hyper network to aggregate local distortion features and global semantic features.
Zhu \textit{et al.} \cite{zhu2021generalizable} propose a meta learning framework for IQA to improve the generalization for unseen image distortions.
Ke \textit{et al.} \cite{ke2021musiq} propose a multi-scale image quality Transformer (MUSIQ), which utilizes the transformer architecture to solve problems of images with varying sizes and aspect ratios.

Due to the insufficient training data, some studies try to learn quality-aware features from ranked image sets, where rank image sets can be generated without laborious labeling. For example, Liu \textit{et al.} \cite{liu2017rankiqa} first train a Siamese Network to learn the quality rank of a pair of images and then fine-tune the trained Siamese Network on existing IQA databases. Ma \textit{et al.} \cite{ma2017dipiq} utilize the RankNet to learn an opinion-unaware BIQA from millions of quality-discriminable image pairs. Other studies utilize quality scores calculated by state-of-the-art FR IQA models \cite{gu2017learning,ye2014beyond} as pseudo labels, so they can generate a large-scale IQA database with pseudo labels to pretrain the IQA model. 
However, most of them are developed on synthetically distorted images and perform poorly on in-the-wild IQA databases, which are demonstrated in Section \ref{results}.

\subsection{Mixed Database Training}
Large-scale annotated databases are very important for training a deep neural network. Since the number of images in existing IQA databases is small, data augmentation is ineluctably adopted to increase training samples in the IQA database. The most widely used method is patch-based training \cite{ye2012unsupervised,po2019novel}. 
Recently, some studies \cite{krasula2019training,li2021unified,zhang2021uncertainty} also use the mixed database training strategy to boost the performance of IQA models. For example, Krasula \textit{et al.} \cite{krasula2019training} train a model by combining multiple video quality databases using a customized cost function. Li \textit{et al.} \cite{li2021unified} propose a unified framework consisting of the relative quality assessor, nonlinear mapping and dataset-specific perceptual scale alignment to train a single video quality assessment model on multiple datasets.  Zhang \textit{et al.} \cite{zhang2021uncertainty} first combine image pairs from the individual database and then train a rank IQA model on image pairs from multiple IQA databases simultaneously. However, their performance is not very promising since they train the whole network on multiple databases and ignore that the function of the regression module is to map the quality-aware features into the quality space of each database. Hence, we argue that the regression modules should be trained on the corresponding target databases.

\section{Proposed Model}
\label{proposed_method}
In this section, we describe the proposed model in detail. A diagram of the network structure is illustrated in Fig. \ref{framework}, which includes two parts, a staircase network for more effective and powerful feature representation and image quality regressors for mapping the quality-aware features to quality score spaces. Then we introduce an iterative mixed database training strategy to train the proposed model on multiple databases simultaneously.
\vspace{-0.5cm}

\subsection{Staircase Network for Feature Extraction}
\label{staircase}
Many successful CNN models such as VGG \cite{simonyan2015very}, GoogleNet \cite{szegedy2015going}, ResNet \cite{he2016deep} follow the same design rule, which gradually reduce the dimension of feature maps and increase the number of feature maps at the same time. This kind of architecture allows the CNN model to learn features from low-level to high-level as the number of network layers deepen, and achieves promising performance in many computer vision tasks such as image recognition \cite{simonyan2015very, szegedy2015going, he2016deep}, object detection \cite{ren2015faster}, etc. However, both the low-level visual features and high-level semantic information play an important role in perceiving the image quality in human visual system (HVS). Moreover, \cite{gao2017deepsim,ali2017image} prove that features extracted from mid-layers and deep-layers are both useful for image quality evaluation. Therefore, it is not optimal to directly use the popular CNN model as the feature extraction module due to the loss of low-level features. Here, we propose the staircase structure to hierarchically integrate the features extracted from intermediate layers, so the model can make full use of features extracted from low-level to high-level visual information.


Generally, the popular CNN architectures can be divided into several stages according to the dimension of feature maps. In each stage, there are several convolutional layers in series to deepen the network. Assume that there are $N_{s}$ stages, and $F_i$ is the feature map extracted from the $i$-th stage, where $i\in [1, 2, ..., N_{s}]$. Since we want to integrate the features extracted from each stage into the final feature representation, a simple method is to fuse the feature maps by element-wise addition operators, i.e.
\begin{eqnarray}
\begin{aligned}
\label{feature_fuse}
& F = \sum^{N_{s}}_{i=1}F_i.
\end{aligned}
\end{eqnarray}
However, there are two problems if we directly use Eq. (\ref{feature_fuse}) as the feature fusion method. 

First, it is observed that the number of channels and the dimension of feature maps in each stage are not the same. Generally speaking, the dimension of the feature maps at the current stage is half that of the previous stage while the number of channels is twice that of the previous stage. So, it is impossible to add the feature maps from different stages directly. In order to make the number of channels and the dimension of feature maps at different stages the same, we introduce a bottleneck structure \cite{he2016deep} consisting of three convolution operations to reduce the dimension and increase the number of channels. The feature map $F_i$ is first reduced to the number of channels to a quarter through the 1$\times$1 convolution layer, which is used to decrease the computation complexities of the following procedures. Then the feature map is reduced to its resolution to half through the 3$\times$3 convolution layer with a stride of 2. Finally, the feature map is passed through the 1$\times$1 convolution layer to increase the number of channels eight times. After that, the feature map $\widetilde F_i$ can be represented as:
\begin{eqnarray}
\begin{aligned}
\label{reduce_d_c}
&  \widetilde F_i = W_{\mathit{1\times1}} W_{\mathit{3\times3}} W_{\mathit{1\times1}} F_i = W F_i ,
\end{aligned}
\end{eqnarray}
where $W_{\mathit{1\times1}}$ and $W_{\mathit{3\times3}}$ are the weight matrices of the 1$\times$1 convolution layer and the 3$\times$3 convolution layer respectively, and $W$ is the product of $W_{\mathit{1\times1}}$, $W_{\mathit{3\times3}}$, and $W_{\mathit{1\times1}}$. Then, we can directly add feature maps from different stages:
\begin{eqnarray}
\begin{aligned}
\label{feature_fuse2}
&  F = \sum^{N_{s}-1}_{i=1} \widetilde F_i  + F_{N_{s}}.
\end{aligned}
\end{eqnarray}

Second, we notice that adding the features from lower layers to the final stage directly will cause the whole network difficult to train. For example, if we use a short connection (include downscaling and channels maps adding operators) to add the features in Stage $1$ to the features in Stage $N_{s}$, it will make the backward propagated gradients tend to pass through the short connection while ignoring the backbone network during training. As a result, it is hard to train the backbone network to extract deep semantic features. Therefore, we propose to hierarchically merge feature maps from different stages to avoid this problem. More specifically, for the feature map $F_1$ from Stage 1, we first downscale its resolution and increase its channels by two convolution layers to obtain $\widetilde F_1$. Then we merge $\widetilde F_1$ with the feature map $F_2$ via element-wise addition and derive $\widetilde F_1^2$. For $\widetilde F_1^2$, we continue to reduce its resolution and increase the channels, and then add it with the feature maps $F_3$ to derive $\widetilde F_1^3$. The same operation is repeatedly implemented until fusing $\widetilde F_1^{N_{s}-2}$ with the feature map $F_{N_{s}-1}$ to derive $\widetilde F_1^{N_{s}-1}$, and $\widetilde F_1^{N_{s}-1}$ is the final feature map extracted from Stage $1$ that needs to be merged into the final feature maps. We then do similar operations for the feature maps from other stages. These procedures can be formulated as:
\begin{eqnarray}
\begin{aligned}
\label{feature_stage1234}
&  \widetilde F_i^{j+1} = W_{ij} \widetilde F_i^j + F_{j+1},
\end{aligned}
\end{eqnarray}
where $i \in [1,2,..,N_{s}-2]$, $j \in [i,...,N_{s}-2]$, and $\widetilde F_i^i = F_i$.

Finally, the quality aware features extracted by the staircase network are represented as:
\begin{eqnarray}
\begin{aligned}
\label{feature_fuse3}
& F = \sum^{N_{s}-2}_{i=1} \widetilde F_i^{N_{s}-1} + W_{N_{s}-1,n}F_{N_{s}-1} + F_{N_{s}} .
\end{aligned}
\end{eqnarray}

\subsection{Image Quality Regressor}
\label{regressor}
After extracting quality-aware features by the staircase network, we need to map these features to the quality scores with a regression model. We first apply the global average pooling (GAP) on the extracted feature maps to produce a feature vector with a dimension of $P \times 1$, where $P$ is the number of final feature maps. Then two Fully Connected (FC) layers are used as the regression model to predict the image quality. The two FC layers consist of 128 and 1 neurons respectively. Finally, we can train the staircase network and image quality regressor in an end-to-end training manner. The Euclidean distance is used as the loss function:
\begin{eqnarray}
\begin{aligned}
\label{loss_function}
& L = \parallel q_{predict} - q_{label} \parallel ^2,
\end{aligned}
\end{eqnarray}
where $q_{predict}$ is the quality score predicted by the proposed model and $q_{label}$ is the ground-truth quality score derived from subjective experiments.

\subsection{Iterative Mixed Database Training Strategy}
\label{mixeddatabase}

\begin{algorithm}[t]
	\caption{ The framework of the iterative mixed database training strategy.}
	\label{alg:training_strategy}
	\begin{algorithmic}[1]
		\REQUIRE ~~\\
		The training IQA database: $D_{t,i}$,\\
		The validation IQA database: $D_{v,i}$,\\
		The number of images in $i$-th training database: $N_i$,\\
		The number of training loops: $L$,\\
		The upper bound number of epochs for $D_{t,i}$  in a loop: $E$,\\
		The criterion function: $C$,\\
		where $i = 1,2,...,m$.
		\ENSURE ~~\\
		The parameters of the proposed model, $\theta_{s,opt}^i$, $\theta_{r,opt}^i$;
		\STATE $\theta_{s,opt}^i \leftarrow \theta_{s}^{0} $
		\FOR{$i \leftarrow 1$; $i<=m$; $i \leftarrow i+1$}
		\STATE $\theta_{r,opt}^i \leftarrow \theta_{r,opt}^0 $
		\ENDFOR
		\FOR{$k\leftarrow1$; $k<=L$; $k\leftarrow k+1$}
		\FOR{$i \leftarrow 1$; $i<=m$; $i \leftarrow i+1$}
		\STATE $E_i = max([N_{max}/N_i],E)$
		\FOR{$j \leftarrow 1$; $j<=E_i$; $j \leftarrow j+1$}
		\STATE Update $\theta_{s}$ and $\theta_{r}^i$ via solving the Eq. (\ref{optimization_fucntion_i}) on the database $D_{t,i}$
		\STATE Validate $\theta_{s}$ and $\theta_{r}^i$ on the database $D_{v,i}$ using criterion function $C$ 		
		\IF{$C(\theta_{s},\theta_{r}^i) > C(\theta_{s,opt}^i,\theta_{r,opt}^i)$ }
		\STATE $\theta_{s,opt}^i = \theta_{s}$, $\theta_{r,opt}^i = \theta_{r}^i$
		\ENDIF
		
		\ENDFOR
		\ENDFOR
		\ENDFOR
		\RETURN $\theta_{s,opt}^i$, $\theta_{r,opt}^i$;
	\end{algorithmic}
	
\end{algorithm}
Deep neural networks usually require large amounts of annotated data to train an effective model. For IQA databases, it is extremely time-consuming and expensive to conduct subjective experiments to obtain the MOS labels of images, which makes the scale of existing IQA databases too small to train a deep CNN model. As a result, increasing the number of training samples is an important way to improve the performance of a deep IQA model. Although many data augmentation strategies such as cropping the patches from original images, rotating images, etc. are utilized to alleviate the problem of insufficient annotated data in previous studies, we argue that it cannot increase the diversity of image content and distortions and thus cannot further improve the generalization abilities of the trained model.

In this section, we propose an iterative mixed database training (IMDT) strategy to train the model on multiple databases simultaneously, so the trained model can benefit from diversities of image content and distortions in each database. 
To solve the problem of the quality scale difference across databases, we only train the feature extraction module on multiple databases, while the image quality regressors are trained on the corresponding databases to map quality-aware features into their quality score spaces. 
Specifically, assume that there are $m$ IQA databases, and the number of images in each database is $N_i$, where $i = 1,2,..,m$. The parameters of the staircase network, image quality regressors, and the whole network are denoted as $\theta_{s}$, $\theta_{r}$, and $\theta$ respectively, and the function of the proposed IQA model is denoted as $f$. Generally, when training the proposed model on a single database, we can use the common gradient descent optimization methods such as Adam \cite{kingma2015adam}, RMSProp, etc. to solve an objective function:
\begin{eqnarray}
\begin{aligned}
\label{optimization_fucntion}
&\theta=  \min_{\theta} \frac{1}{T} \sum_{j=1}^{T} \parallel f(I_j;\theta) - q_{label,j} \parallel ^2,
\end{aligned}
\end{eqnarray}
where $T$ is the total number of training images in a batch. The parameters $\theta$ are updated after each batch in order to decrease the loss of the objective function. Finally, we will choose the parameters with the minimum loss as the best trained model. 

When training the proposed model on multiple databases, we can consider this procedure as solving multiple sub-problems, where the result of each sub-problem corresponds to the best trained model on one database. Since only the feature extraction module needs to be trained on multiple databases, we add $m$ image quality regressors after the staircase network, which are trained on their corresponding databases respectively. Accordingly, the objective functions of $m$ sub-problems are formulated as:
\begin{eqnarray}
\begin{aligned}
\label{optimization_fucntion_i}
&\theta_s, \theta_r^i =  \min_{\theta_s, \theta_r^i} \frac{1}{T} \sum_{j=1}^{T} \parallel f(I;\theta_{s},\theta_{r}^i) - q_{label,j}^i \parallel ^2,
\end{aligned}
\vspace{-0.2cm}
\end{eqnarray}
where $i = 1,2,...,m$. Then we alternatively solve $m$ sub-problems through using gradient descent method to reduce the loss on the corresponding databases. The relationship between these sub-problems is that the parameters of the feature extraction module $\theta_{s}$ are updated across all sub-problems. In particular, we first initialize the parameters $\theta_{s}$ and $\theta_{r}^i$ with $\theta_{s}^0$ and $\theta_{r}^{i,0}$. 
We define that a \textbf{loop} is that $m$ sub-problems have been solved in one epoch.
For the $d$-th loop, we use the gradient descent method to solve the objective function Eq. (\ref{optimization_fucntion_i}) to obtain $\theta_{s}^{m(d-1)+i}$ and $\theta_{r}^{i,d}$ on the $i$-th database in one epoch. Then $\theta_{s}^{m(d-1)+i}$ is used to initialize $\theta_{s}^{m(d-1)+i+1}$, and we solve the objective function Eq. (\ref{optimization_fucntion_i}) to obtain $\theta_{s}^{m(d-1)+i + 1}$ and $\theta_{r}^{i+1,d}$ on the ($i+1$)-th database. Note that the number of images in different databases may vary greatly, so it is unbalanced to update the parameters if we train all databases for one epoch in a loop. Therefore, we set the number of epochs of a database trained in a loop as $max([N_{max}/N_i],E)$, where $N_{max}$ is the maximum number of images in IQA databases, $[\cdot]$ is the rounding operator, and $E$ is the upper bound of epoch number in a loop. We have summarized the framework of iterative mixed database training strategy in Algorithm \ref{alg:training_strategy}.

\section{Experimental Results}
\label{results}
In this section, we first present the experimental protocol in detail and then report the comparison results between the proposed model and other state-of-the-art BIQA models on six in-the-wild IQA databases and four synthetic IQA databases. The ablation studies are conducted to validate the effectiveness of each module in the proposed model.

\subsection{Experimental Protocol}
\label{protocol}
\subsubsection{Test Databases}The proposed method is mainly validated on the following six authentically distorted IQA databases:
\begin{itemize}
	\item CLIVE: LIVE In the Wild Image Quality Challenge Database (CLIVE) \cite{ghadiyaram2016massive} consists of 1,162 images with diverse authentic distortions captured by various mobile devices.
	
	\item BID: BID \cite{ciancio2011no} is a blur image database that contains 586 images with realistic blur distortion such as out-of-focus, simple motion, complex motion blur, etc. The images were shot under various lighting conditions and exposure times.
	
	\item KonIQ10K: KonIQ10K \cite{hosu2020koniq} contains 10,073 images which are selected from the large public multimedia database YFCC100m. The selected images cover a wide and uniform range of distortions in terms of quality indicators such as brightness, colorfulness, contrast, noise, sharpness, etc.
	
	\item SPAQ: Smartphone Photography Attribute and Quality (SPAQ) \cite{fang2020perceptual} database consists of 11,125 images taken by 66 kinds of mobile devices. SPAQ database covers a large range of scene categories such as animal, human, plant, indoor scene, cityscape, landscape, night scene, etc. Besides providing image quality scores, the SPAQ database also gives image attribute scores (contrast, brightness, noisiness, colorfulness, and sharpness). In this paper, we only focus on the overall image quality.
	
	\item FLVIE: FLIVE\cite{ying2020from} is the largest in-the-wild IQA database by far, which contains about 40,000 real-world distorted images and 120,000 randomly cropped patches. The images are selected from five public image databases (including AVA\cite{murray2012ava}, VOC\cite{everingham2010the}, EMOTIC\cite{kosti2017emotic}, and CERTH Blur \cite{mavridaki2014no}) with diverse contents, sizes, and aspect ratios. In this paper, we validate the proposed model on both FLIVE and FLIVE Patch databases.
\end{itemize}


Besides the in-the-wild IQA databases, we also validate the proposed model on four synthetic IQA databases, including:

\begin{itemize}
	\item LIVE: LIVE \cite{sheikh2005live} is the most commonly used synthetic IQA database. It includes 770 lossy images derived from 29 pristine images by degrading them with five different types of distortions: additive white Gaussian noise, Gaussian blur, JPEG compression, JPEG2000 compression, and fast fading.
	\item CSIQ: CSIQ \cite{larson2010most} contains 886 distorted images created from 30 original images. Six types of distortions are considered in the CSIQ database, which are Gaussian blur, additive white Gaussian noise, JPEG compression, JPEG2000 compression, global contrast decrements, and additive pink Gaussian noise at four or five distortion levels respectively.
	\item Kadid10K: Kadid10k \cite{lin2019kadid} is the largest synthetic IQA image database so far, which includes 10,125 images degraded from 81 pristine images. Each pristine image are degraded by 25 distortions in 5 levels. More details about distortion types in Kadid10k can be referred to \cite{lin2019kadid}.
	\item LIVEMD: LIVEMD \cite{jayaraman2012objective} is a multiply distorted IQA database. It contains 15 reference images and 450 distorted images. Two multiply distortion types are consider in the LIVEMD database, which are blur followed by JPEG compression and blur followed by noise respectively.
\end{itemize}

The subjective experiments of CLIVE, KonIQ-10K, FLIVE, FLIVE Patch, and Kadid10k databases were conducted on the online crowdsourcing system, while BID, SPAQ, LIVE, CSIQ, and LIVEMD databases were evaluated in a well-controlled laboratory environment. Note that the image content in these above databases does not overlap, so there is no risk of accessing some images that may be in the training set of one dataset and meanwhile in the testing set of another dataset.

\begin{table*}
	\centering
	\renewcommand{\arraystretch}{1.25}
	\caption{Performance of eleven state-of-the-art methods and the proposed model on six in-the-wild IQA databases. The best performing models in IQA categories are highlighted in each column. The $^\ast$ means that the results are cited from the original paper.}
	\label{performance}
		\begin{tabular}{c|cc|cc|cc|cc|cc|cc}
			\toprule[.15em]
			Database & \multicolumn{2}{c|}{CLIVE} & \multicolumn{2}{c|}{BID} & \multicolumn{2}{c|}{KonIQ10k} & \multicolumn{2}{c|}{SPQA} & \multicolumn{2}{c|}{FLIVE} & \multicolumn{2}{c}{FLIVE Patch}  \\
			
			Criterion & SRCC &  PLCC & SRCC & PLCC & SRCC & PLCC & SRCC & PLCC & SRCC & PLCC & SRCC & PLCC    \\
			\hline
			
			QAC \cite{xue2013learning}   & 0.0686	& 0.0138&0.3258	&  0.3229 &	0.3430&0.2961	& 0.0465	&0.1072	&0.1042	& 	0.0656& 0.1720	& 	0.1042  \\
			NIQE \cite{mittal2012making}  &	0.4536&	0.4676& 0.4772	&0.4713	&	0.5260 &	0.4745&0.6973	& 0.685	& 0.1048	& 0.1409	&0.3211	&0.2826	  \\

			ILNIQE \cite{zhang2015feature}   & 0.4531	&0.5114	& 0.4946	&  0.4538	& 0.5029	& 0.4956	&	0.7194&	0.654& 0.2188	& 	0.2547& 0.5306	&	0.5231 \\			
			BRISQUE \cite{mittal2012no}&  0.6005 & 0.6211	& 0.5736	& 0.5401	& 0.715	&0.7016	&0.8021	&0.8056	&0.3201	& 0.3561 &0.5372&	0.5843\\
			BMPRI \cite{min2018blind} &	0.4868& 0.5229	& 0.5154	&0.4583	&0.6577	&  0.6546  &0.7501	&0.7544	&	0.2737&	0.3146&0.5839	&0.6142	\\
			\hline
			CNNIQA \cite{kang2014convolutional}   &0.6269 	& 0.6008  & 0.6163	& 0.6144	& 0.6852	& 0.6837	& 0.7959 & 0.7988 & 0.3059 	& 0.2850 & 0.6005 	& 0.5379 	  \\
			
			WaDIQaM-NR \cite{bosse2017deep} &0.6916 &0.7304	& 0.6526	&0.6359	& 0.7294	&0.7538	& 0.8397 	& 0.8449 	& 0.4346 	& 0.4303	& 0.6995	& 0.7197 \\
			
			SFA \cite{li2018has}    & 0.8037&0.8213&0.8202&0.8253&0.8882&0.8966&0.9057	& 0.9069	& 0.5415&0.626&0.7175&0.7501\\
			DB-CNN  \cite{zhang2020blind}   &0.8443 	& 0.8624	&0.8450 	& 0.8590	&0.8780 	&0.8867 	&0.9099&0.9133& 0.5537	&0.6518 &0.7509&0.7869 \\
			HyperIQA \cite{su2020blindly}    &0.8546 &0.8709	&0.8544	&  0.8585	&0.9075	&0.9205	&0.9155&0.9188 &0.5354	&0.6228	&0.7489&0.7850 \\
			UNIQUE \cite{zhang2021uncertainty} &0.854$^\ast$&0.890$^\ast$&0.858$^\ast$&0.873$^\ast$&0.896$^\ast$&0.901$^\ast$&-&-&-&-&-&-\\
			Proposed     & \textbf{0.8992} 	&\textbf{0.9175} 	&\textbf{0.9128} 	&\textbf{0.9284} 	&\textbf{0.9209} 	& \textbf{0.9362} 	&\textbf{0.9238} 	& \textbf{0.9273} &	\textbf{0.5821} &\textbf{0.6936} 	& \textbf{0.7679} 	&\textbf{0.8012} \\
			\bottomrule[.15em]
	\end{tabular}
\vspace{-0.2cm}
\end{table*}

\subsubsection{Implementation Details}

We use ResNet50 as the backbone of the staircase network. Since there are six databases that need to be validated, we add six image quality regressors after the staircase network. The weights of the backbone are initialized by training on ImageNet, and other weights are randomly initialized. For the FLIVE database, we use the same pre-processing method in \cite{ying2020from} to white fill images to the resolution of 340$\times$340. Since the resolution of images in the FLIVE Patch database is very small and is usually less than 256, we then white fill images in the FLIVE Patch database to the resolution of 256$\times$256. For images in other databases, we resize the resolution of the minimum dimension of images as 380 while maintaining their aspect ratios. In the training stage, the input images in the FLIVE Patch database and other databases are randomly cropped with resolutions of 224$\times$224 and 320$\times$320 respectively, and in the testing stage, we crop the four corners and center patch with the same resolution of 224$\times$224 for images in the FLIVE Patch database and 320$\times$320 for images in other databases. The quality score of each testing image is averaged by the predictive scores of five patches. The proposed model is implemented with PyTorch. The Adam optimizer with the initial learning rate 0.00001 and training batch size 30 are used for training the proposed model on a server with NVIDIA GTX 2080Ti. The hyper parameters $L$ and $E$ are set as 3 and 20 respectively.

\subsubsection{Comparing Algorithms}
We compare the proposed models with the state-of-the-art BIQA models, including:
\begin{itemize}
	\item Handcrafted feature based BIQA models: QAC\cite{xue2013learning}, NIQE\cite{mittal2012making}, ILNIQE\cite{zhang2015feature}, BRISQUE\cite{mittal2012no}, and BMPRI\cite{min2018blind}. Among them, QAC, NIQE, and ILNIQE are OU-BIQA models, which do not need to be trained on the training set.
	\item Deep learning based BIQA models: CNNIQA \cite{kang2014convolutional}, WaDIQaM-NR \cite{bosse2017deep}, SFA \cite{li2018has}, DB-CNN \cite{zhang2020blind}, HyperIQA \cite{su2020blindly}, and UNIQUE \cite{zhang2021uncertainty}. Note that UNIQUE is also trained on multiple databases.
\end{itemize}
Except UNIQUE\footnote[1]{The training method of UNIQUE requires the standard deviation value of each MOS, which is not provided by the SPQA, FLIVE, and FLIVE Patch databases. Therefore, we directly list the results of UNIQUE in the original paper \cite{zhang2021uncertainty}.}, we retrained the other compared models on the six IQA databases for a fair comparison. For OU-BIQA models, the performance is validated on the testing databases directly.

\subsubsection{Evaluation Criteria}Two common criteria are adopted to evaluate the performance of IQA models, which are Pearson linear correlation coefficient (PLCC) and Spearman rank-order correlation coefficient (SRCC). 
These two criteria have different meanings for demonstrating the performance of IQA models.
To be more specific, PLCC reflects the prediction linearity of the IQA algorithm and SRCC indicates the prediction monotonicity. An excellent IQA model should obtain the value of SRCC and PLCC close to 1.

All databases are split into the training set with 80\% distorted images and the test set with 20\% distorted images. For synthetically distorted images, the distorted images corresponding to the same reference image are assigned to the same set to ensure complete separation of the training and testing content. We randomly split the databases for 10 times, and report the median values of SRCC and PLCC.

\subsection{Performance Comparison with the State-of-the-art Methods}
\begin{figure*}[htbp]
	\centering
	\subfigure[LIVEC]
	{\label{fig1.1}
		\includegraphics[width=0.28\textwidth]{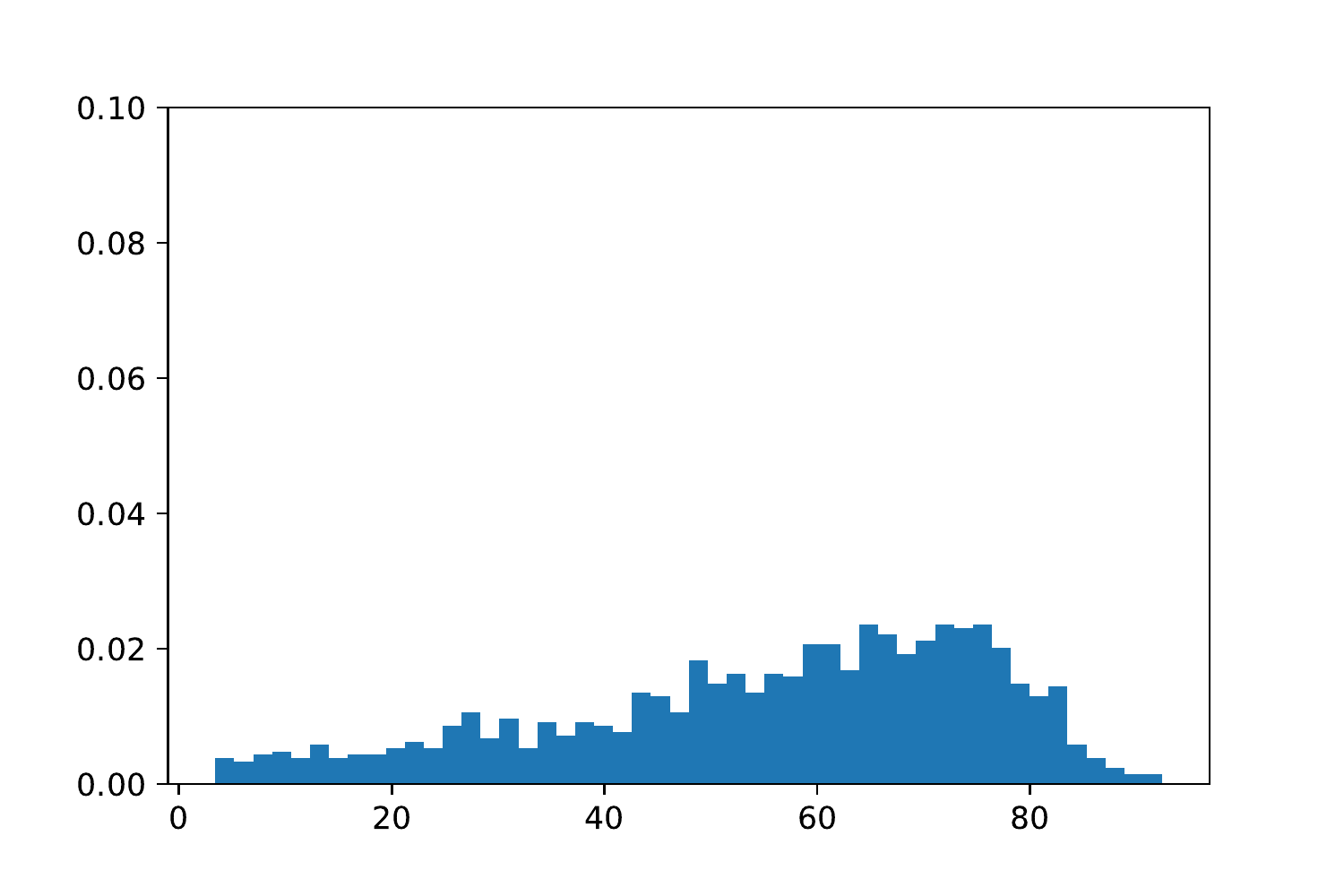}
	}
	\subfigure[BID ]
	{\label{fig1.2}
		\includegraphics[width=0.28\textwidth]{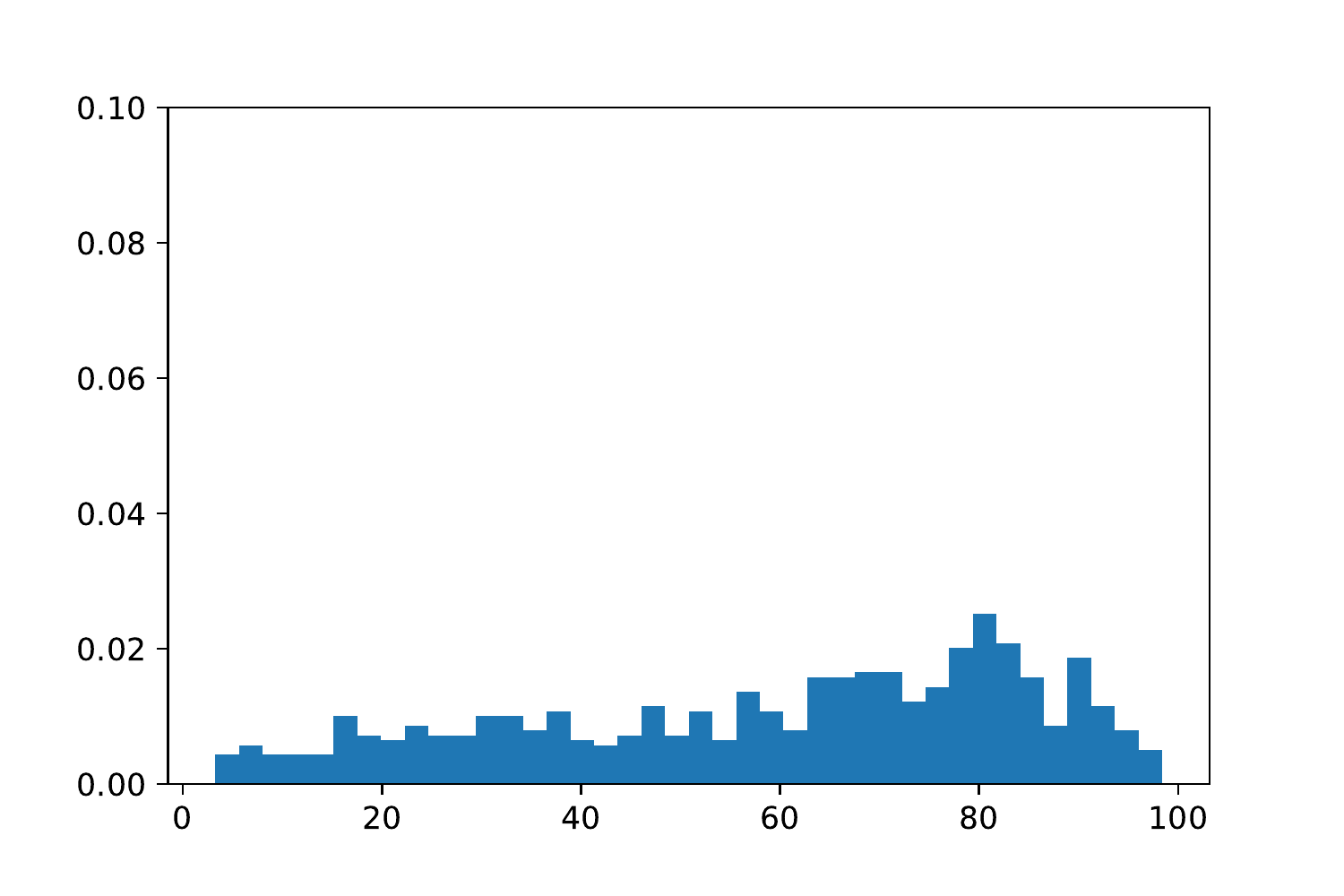}
	}
	\subfigure[KonIQ10k ]
	{\label{fig1.3}
		\includegraphics[width=0.28\textwidth]{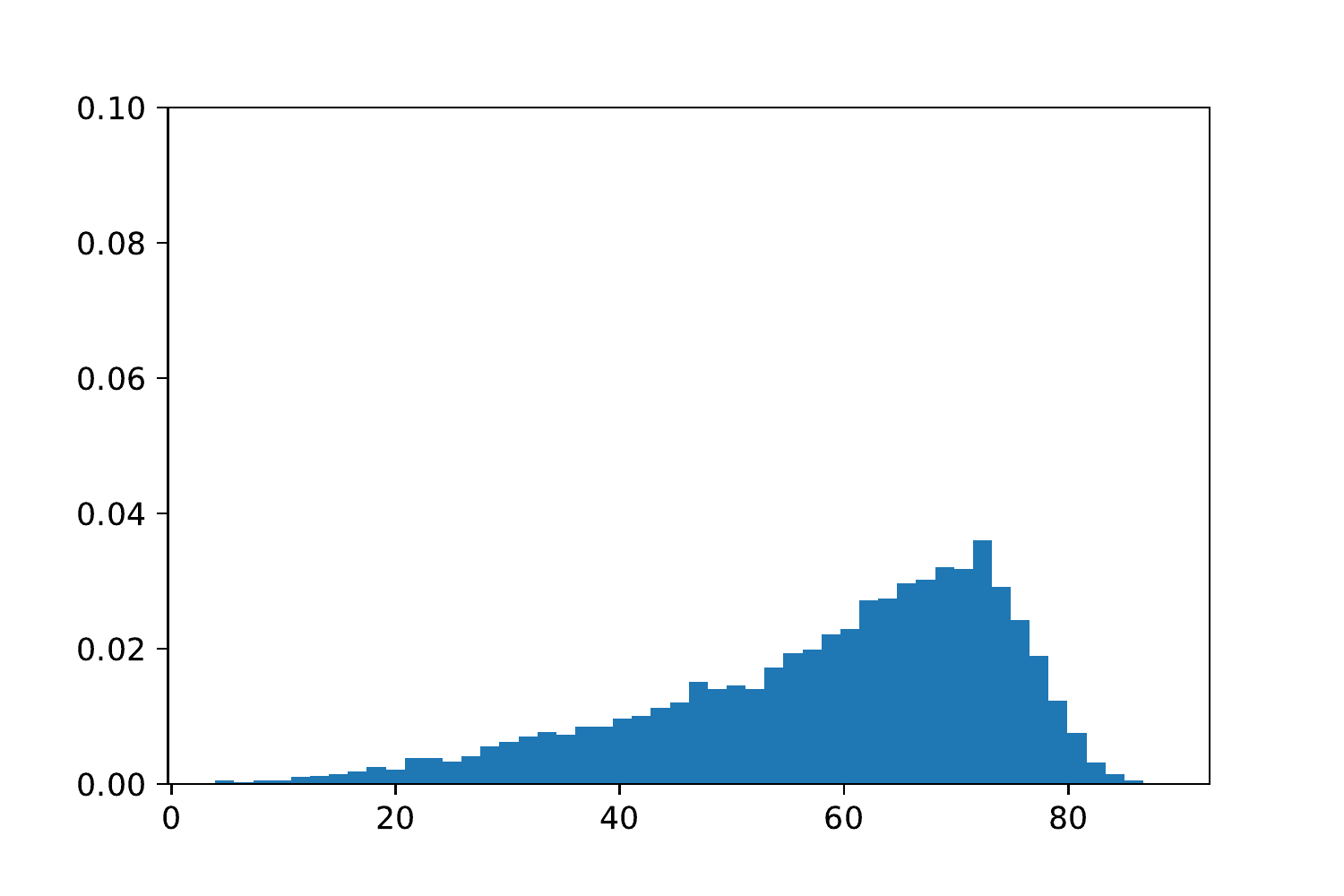}
	}
	\subfigure[SPAQ ]
	{\label{fig1.4}
		\includegraphics[width=0.28\textwidth]{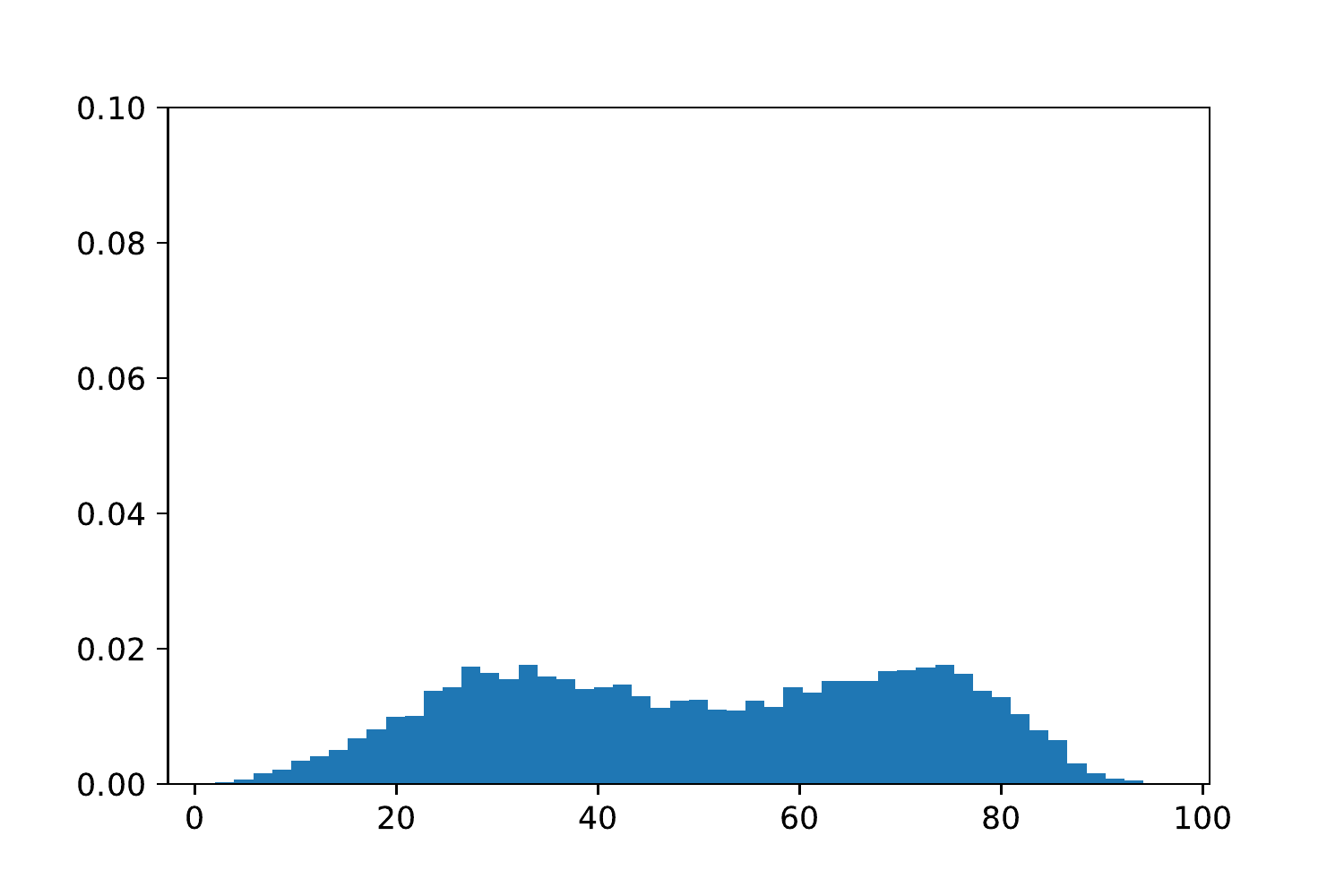}
	}
	\subfigure[FLIVE ]
	{\label{fig1.5}
		\includegraphics[width=0.28\textwidth]{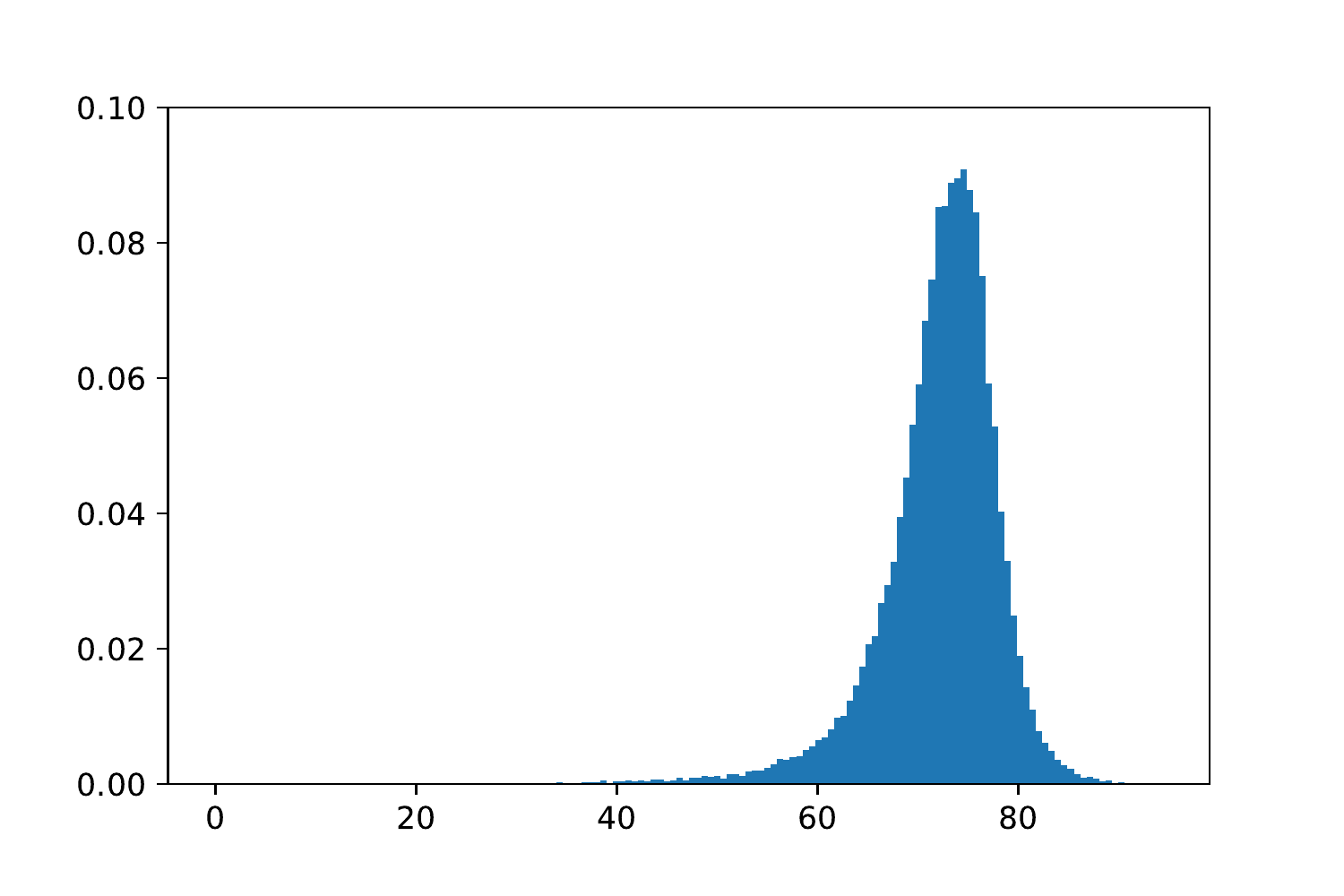}
	}
	\subfigure[FLIVE Patch ]
	{\label{fig1.6}
		\includegraphics[width=0.28\textwidth]{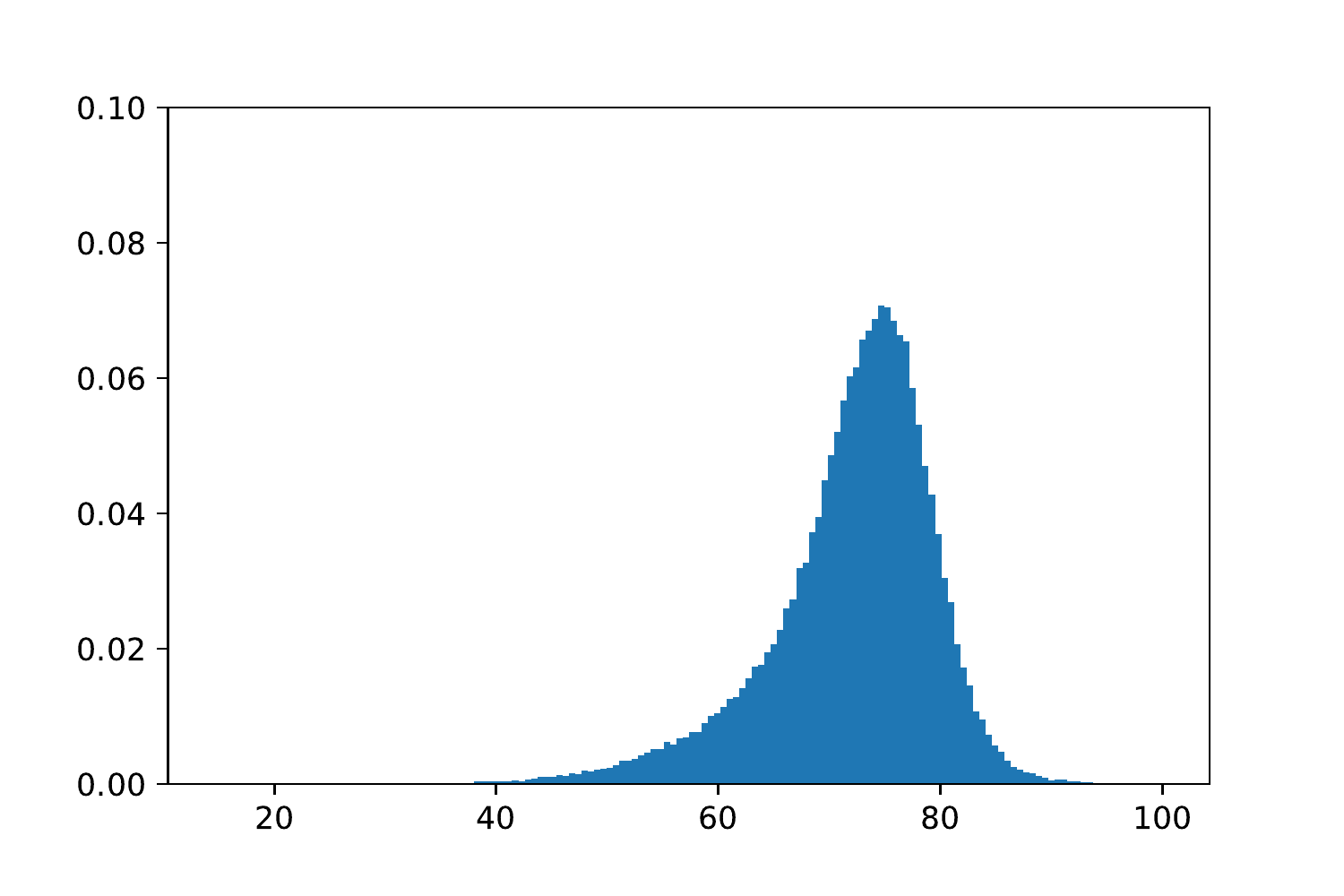}
	}
	
	\caption{MOS histograms of six in-the-wild databases.}
	\label{fig:1}
\end{figure*}

\subsubsection{In-the-wild Databases}
\label{in-the-wild databases}
The performance results on the in-the-wild databases are summarized in Table \ref{performance}. From Table \ref{performance}, we first observe that the proposed model achieves the best performance on all six in-the-wild IQA databases and it leads by a significant margin, which indicates that the proposed model has more powerful representation abilities for the quality of in-the-wild images than other deep learning based methods as well as handcrafted feature based methods. We also notice that the proposed model significantly improves the performance of CLIVE and BID databases. Note that the number of images in CLIVE and BID databases is far less than the other four databases, so it is difficult for these deep learning models to learn a better feature representation if we train the model on CLIVE and BID databases individually.
Fortunately, with the help of the IMDT strategy, the proposed model can learn the feature representation from more diverse image content and distortions in other databases, and it boosts the performance of the proposed model on BID and CLIVE databases to the same level as other large-scale databases such as KonIQ10k and SPAQ. 

Then all handcrafted feature based models perform poorly on in-the-wild IQA databases, and their performance is obviously lower than deep learning based models, which reflects that handcrafted features are difficult to model the quality of images captured in various in-the-wild environments. 
Third, other deep learning based models such as HyperIQA and SFA also use ResNet50 as the backbone for extracting features and UNIQUE is trained on multiple databases, but their performance is all inferior to the proposed model, which indicates the superiority of the staircase structure and the iterative mixed database training strategy for improving the representation ability of the model.

Finally, we observe that the performance of all models on the FLIVE database is far behind the other five databases, though the FLIVE database is the largest in-the-wild IQA database to date. To investigate the reason for this phenomenon, we illustrate the histogram of mean opinion scores (MOS) of six in-the-wild databases in Fig. \ref{fig:1}. From Fig. \ref{fig:1}, it is seen that the MOS distribution of the FLIVE database is mainly concentrated on a high-quality score (MOS of about 75), and its shape is much narrower and peakier than other databases, while the MOS distribution of other databases covers more evenly from low-quality scores to high-quality scores. So, it is reasonable to infer that most images in the FLIVE database have similar and high-quality scores and it is difficult to distinguish their relative quality scores. As a result, it requires the BIQA model can evaluate the image quality in a \textbf{fine-grained way} \cite{zhang2021fine}, which is very challenging even for deep learning based models. With the rapid development of camera devices and computational photography technologies, it is noted that more and more images shared on social media applications are high-quality. Therefore, it is necessary to develop BIQA models that can evaluate image quality in a fine-grained way, which is also our future research direction. 

\begin{figure*}[htbp]
	\begin{center}
		\subfigure[CLIVE]{\includegraphics[width=4.5cm]{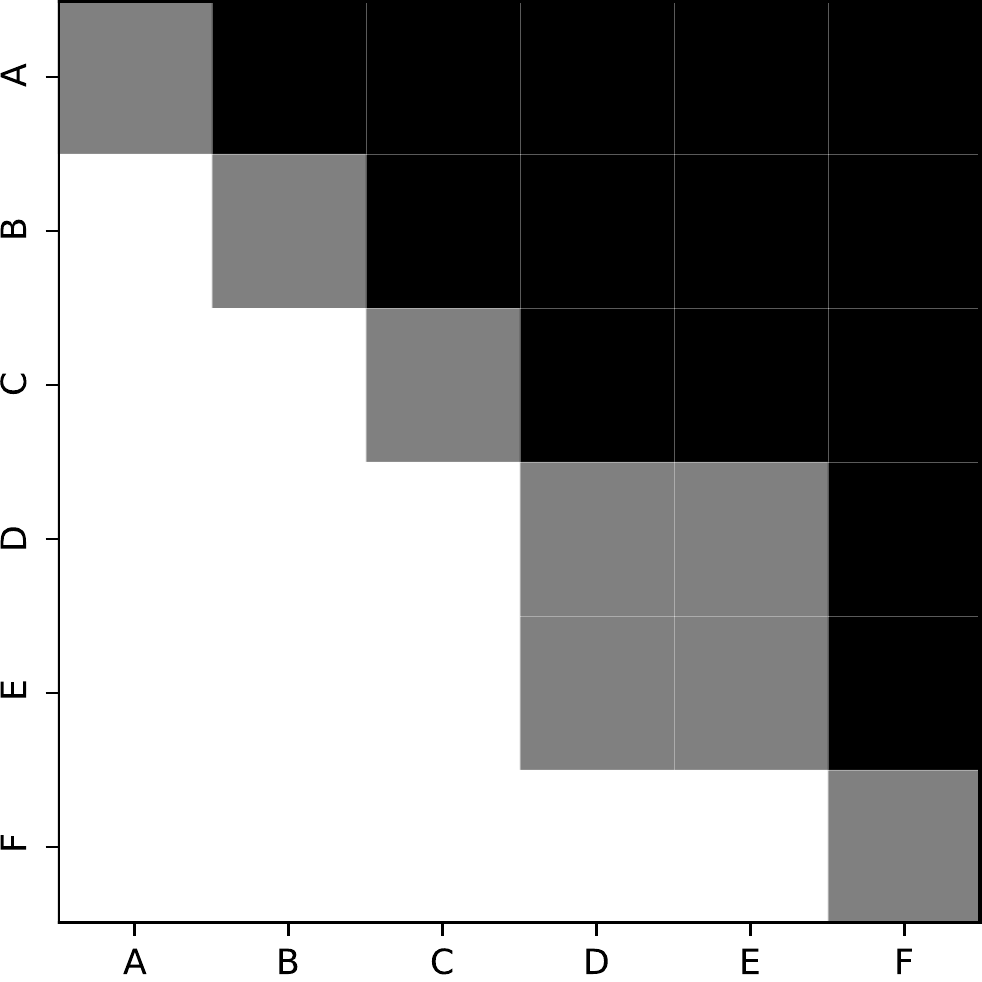}}
		\subfigure[BID]{\includegraphics[width=4.5cm]{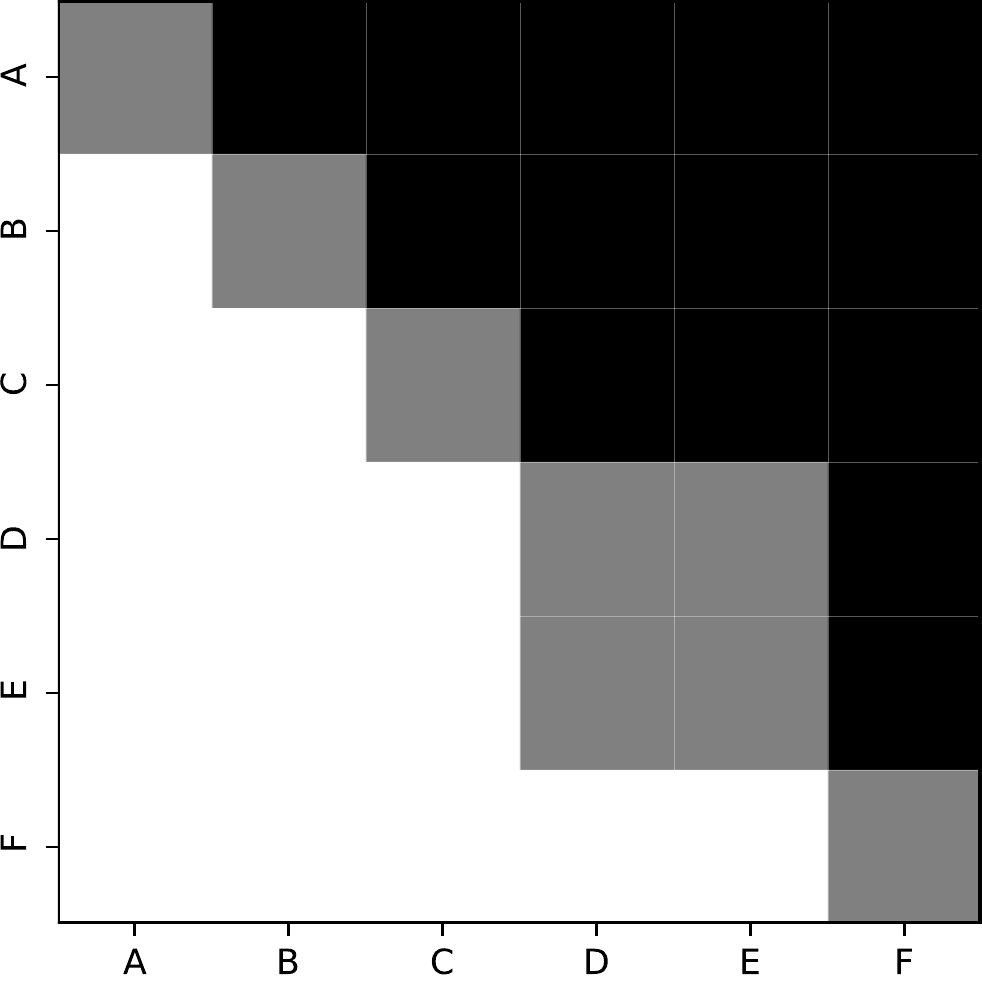}}
		\subfigure[KonIQ10k]{\includegraphics[width=4.5cm]{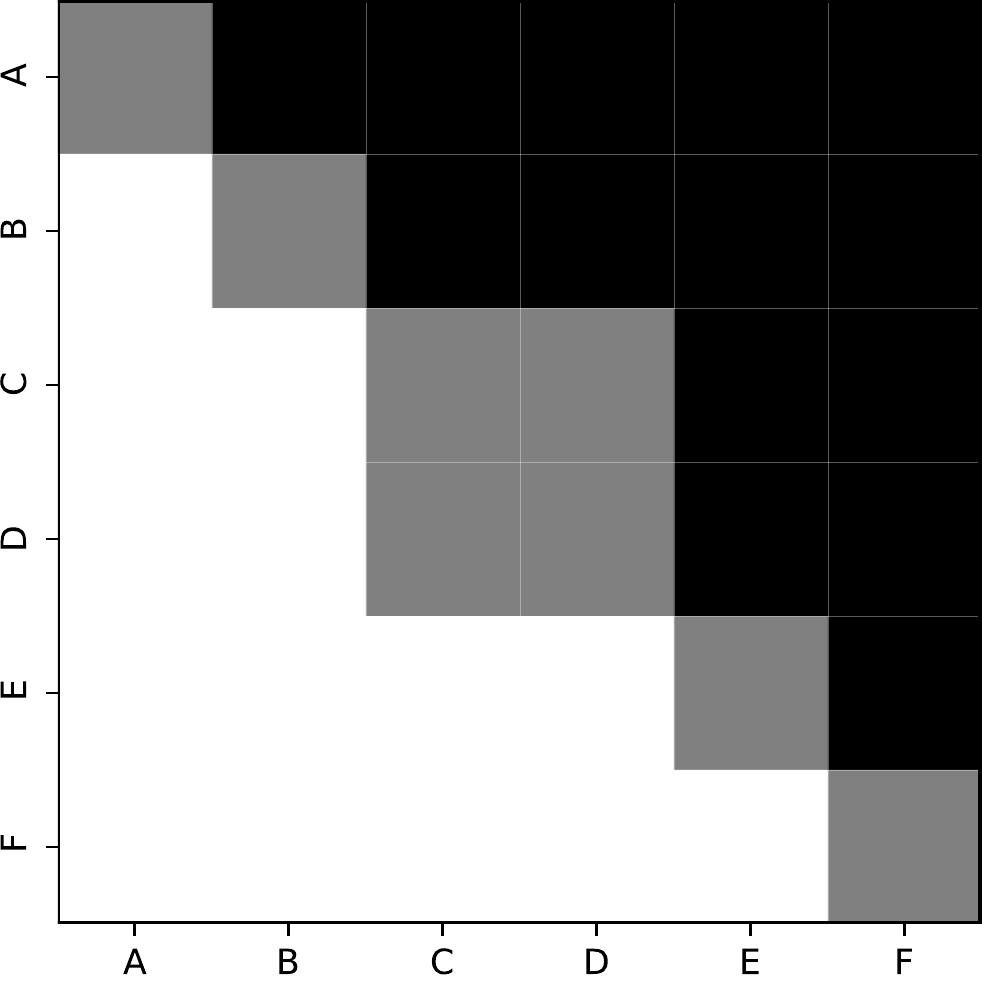}}
  \subfigure[SPQA]{\includegraphics[width=4.5cm]{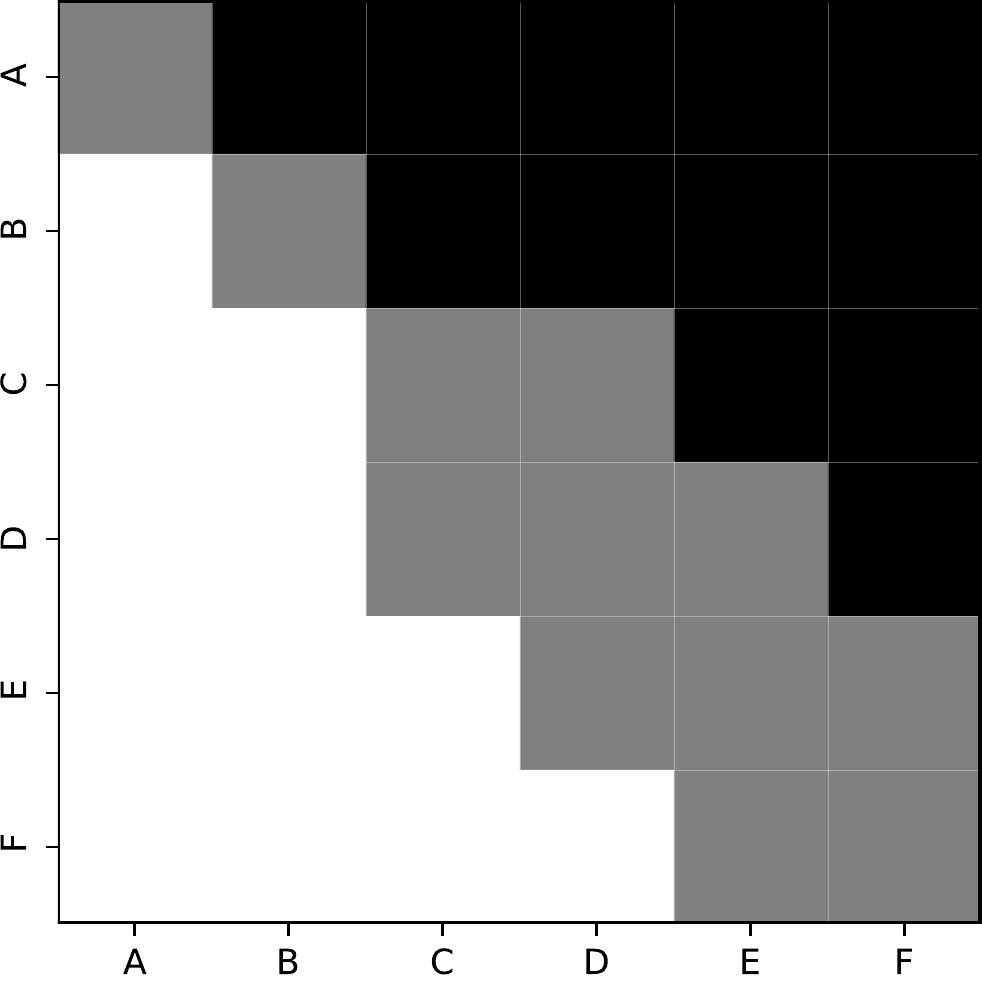}}
  \subfigure[FLIVE]{\includegraphics[width=4.5cm]{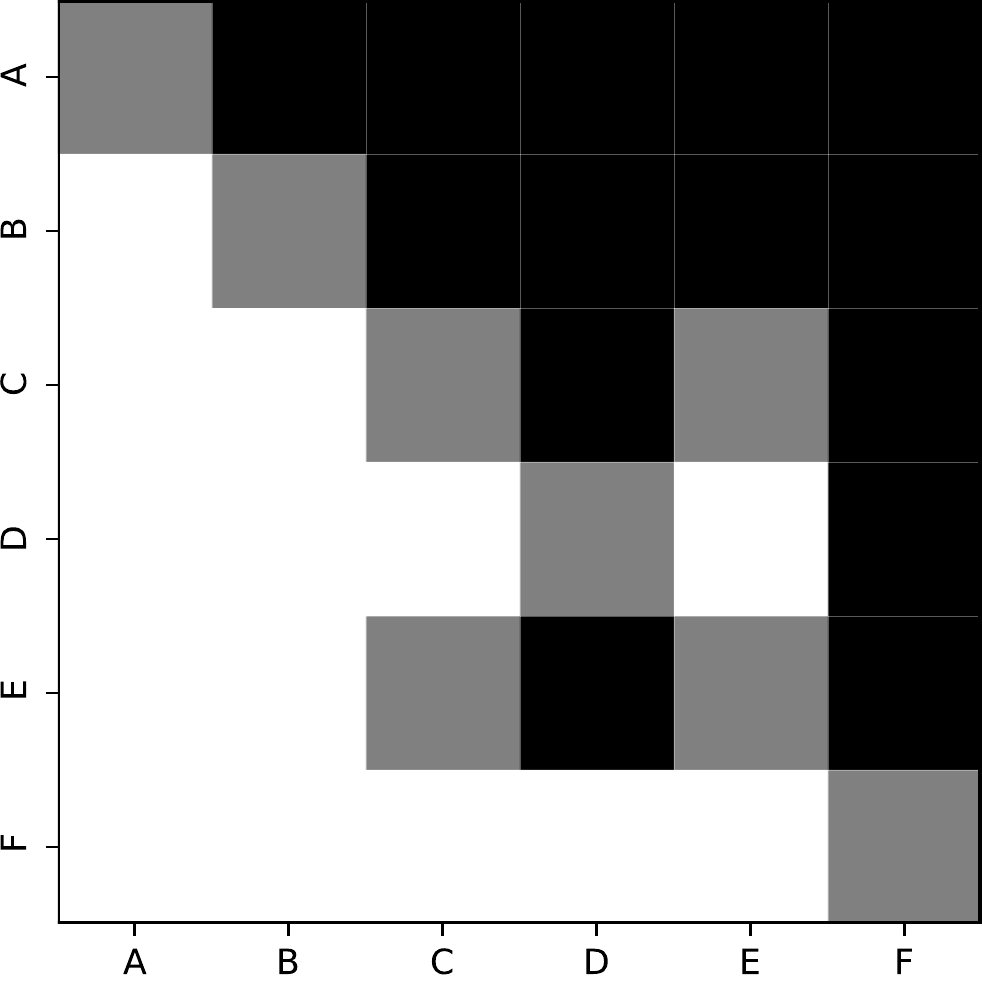}}
  \subfigure[FLIVE Patch]{\includegraphics[width=4.5cm]{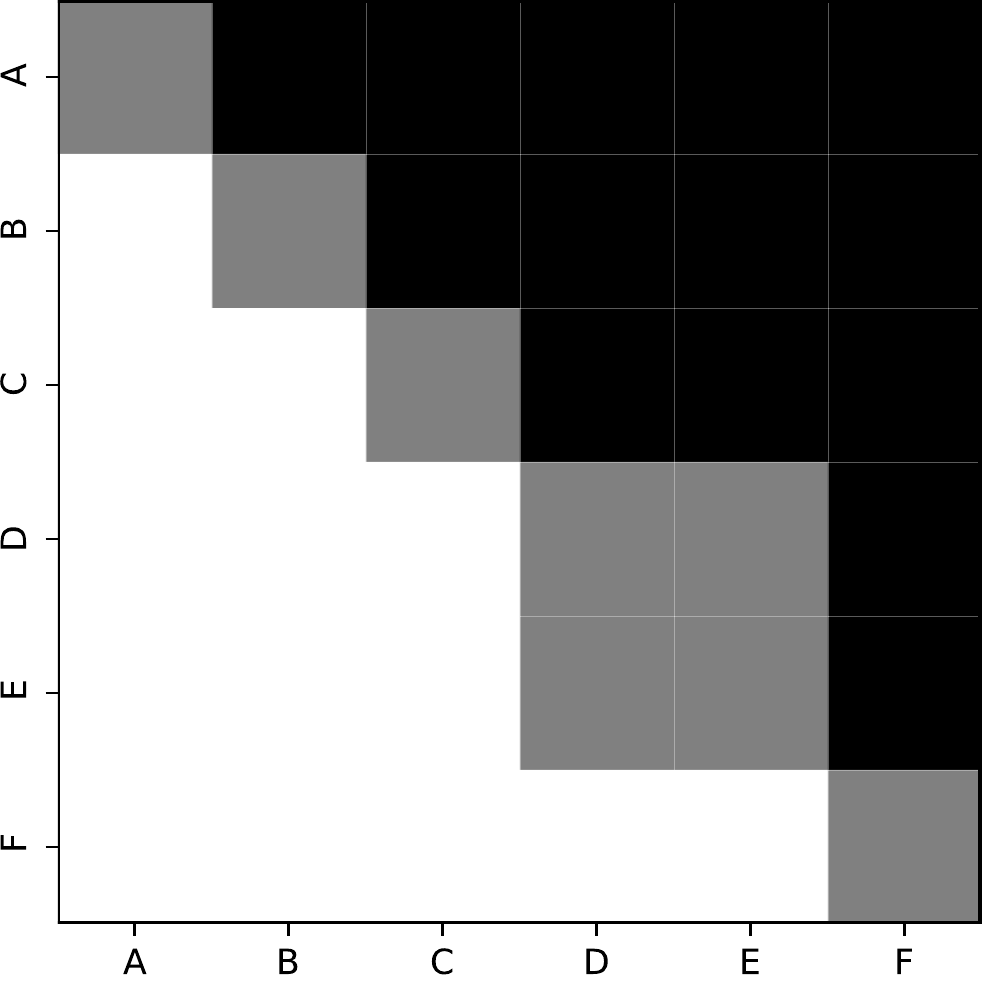}}
	\end{center}
	\caption{Statistical significance comparison between the proposed model and other deep learning-based IQA methods on six authentically distorted IQA databases. A black/white block $(i,j)$ means the method at row $i$ is statistically worse/better than the one at column $j$. A gray block $(m,n)$ means the method at row $m$ and the method at $n$ are statistically indistinguishable. The IQA methods denoted by A-F are CNNIQA, WaDIQaM-NR, SFA, DB-CNN, HyperIQA, UNIQUE, and the proposed method respectively.}
	\label{fig:statistical_significance_authentic}
\end{figure*}

To further analyze the performance of the proposed model and other deep learning based IQA model, we conduct the statistical significance test in \cite{sheikh2006statistical} to measure the difference between the evaluated quality scores and the subjective ratings. We present the results of the statistical significance test on six authentically distorted IQA databases in Fig. \ref{fig:statistical_significance_authentic}. From Fig. \ref{fig:statistical_significance_authentic}, we can observe that the proposed method is statistically superior to other deep learning based IQA methods, which further demonstrates the effectiveness of our model.

\subsubsection{Synthetically Distorted Databases} We list the performance of the proposed model and the compared BIQA methods on four synthetic databases in Table \ref{performance_synthetic_distortion_SRCC} and Table \ref{performance_synthetic_distortion_PLCC}. From Table \ref{performance_synthetic_distortion_SRCC} and Table \ref{performance_synthetic_distortion_PLCC}, it is observed that the proposed model achieves competitive performance on four synthetic IQA databases compared with other state-of-the-art models. Specifically, our model outperforms all hand-crafted feature based methods on these four synthetic IQA databases. For deep learning based models, our model achieves similar performance to DB-CNN and HyperIQA, which are the best performing IQA models on synthetic IQA databases so far, and outperforms other deep learning methods. It should note that most compared models are developed on synthetic IQA databases, while our model does add any modules specifically designed for synthetic distortions, so the comprising performance also demonstrates that the proposed model has a strong generalization ability on quality assessment of other image distortion types.

We also notice that the IMDT strategy does not significantly improve the performance of small-scale databases such as LIVE, CSIQ, and LIVEMD. The reason is that the quality of synthetically distorted images is more related to the distortion type and degree than the image content. Note that there are only five, six, and two kinds of distortion types in the LIVE, CSIQ, and LIVEMD databases respectively. Hence, the number of images in these databases may be enough to train an effective deep model to learn a good representation for these synthetic distortions.

\begin{table}
	\footnotesize
	\centering
	\renewcommand{\arraystretch}{1.25}
	\caption{SRCC of ten state-of-the-art methods and the proposed model on four synthetically distorted IQA databases. The three best performing models are highlighted in each column. The $^\ast$ means that the results are cited from the original paper.}
	\label{performance_synthetic_distortion_SRCC}
	\begin{tabular}{c|cccc}
		
		\toprule[.15em]
		Database & LIVE & CSIQ & Kadid10k & LIVEMD   \\
		\hline
		
		QAC \cite{xue2013learning}   &0.8796&0.4856  & 0.2383  &0.3588  \\
		
		NIQE \cite{mittal2012making} &0.9115  &0.6402  &0.3670  &0.6025  \\
		
		ILNIQE \cite{zhang2015feature}    &0.9028 &0.8238  & 0.5516 &0.9035    \\
		BRISQUE \cite{mittal2012no} &0.9423 &0.6983  &0.5279  &0.8915   \\
		BMPRI \cite{min2018blind} &0.9284  	&0.7713 &0.5604 &0.8308  \\
		\hline
		CNNIQA \cite{kang2014convolutional} &0.9481 	&0.8834  &0.6539  &0.8850  \\
		WaDIQaM-NR \cite{bosse2017deep} &0.9110 &0.8481 &\textbf{0.8773}  &0.8551 \\
		SFA \cite{li2018has} &0.9332 &0.8266 &0.7025 &0.8396 \\
		DB-CNN \cite{zhang2020blind} &\textbf{0.9659} 	&\textbf{0.9254} &\textbf{0.8681} &\textbf{0.9253} \\
		HyperIQA \cite{su2020blindly} & 0.9558 	&\textbf{0.9229}& 0.8391 & \textbf{0.9291}  \\
		UNIQUE \cite{zhang2021uncertainty} & \textbf{0.969$^\ast$} 	&0.902$^\ast$& \textbf{0.878$^\ast$} & \textbf{-}  \\
		Proposed    & \textbf{0.9655} 	&\textbf{0.9190} 	&0.8663 	&\textbf{0.9248} \\
		
		\bottomrule[.15em]
		
	\end{tabular}
\end{table}

\begin{table}
	\footnotesize
	\centering
	\renewcommand{\arraystretch}{1.25}
	\caption{PLCC of ten state-of-the-art methods and the proposed model on four synthetically distorted IQA databases. The three best performing models are highlighted in each column. The $^\ast$ means that the results are cited from the original paper.}
	\label{performance_synthetic_distortion_PLCC}
	\begin{tabular}{c|cccc}
		\toprule[.15em]
		
		Database & LIVE & CSIQ & Kadid10k & LIVEMD   \\
		\hline
		
		QAC \cite{xue2013learning}   &0.8770   &0.7070  &0.4028   &0.4363    \\
		
		NIQE \cite{mittal2012making}  &0.6511&0.7330  & 0.4457   &0.6887   \\
		
		ILNIQE \cite{zhang2015feature}    &0.7067  &0.8642   &0.5938  &0.8769   \\
		BRISQUE \cite{mittal2012no} & 0.9423 &0.7507  &0.5590  &0.9130   \\
		BMPRI \cite{min2018blind} & 0.9291 	&0.8218  &0.6090  &0.8823   \\
		\hline
		CNNIQA \cite{kang2014convolutional} & 0.9467	& 0.8969  & 0.6772  & 0.9165 \\
		WaDIQaM-NR \cite{bosse2017deep} & 0.9226	&0.8796 & \textbf{0.8852} & 0.8570\\
		SFA \cite{li2018has} & 0.9346	& 0.8475 & 0.6986 & 0.7956 \\
		DB-CNN \cite{zhang2020blind} & \textbf{0.9655}	&\textbf{0.9464} &0.8721 & \textbf{0.9284}\\
		HyperIQA \cite{su2020blindly} & 0.9625 &\textbf{0.9593} &0.8496&\textbf{0.9362}\\
		UNIQUE \cite{zhang2021uncertainty} & \textbf{0.968$^\ast$} 	&0.927$^\ast$& \textbf{0.876$^\ast$} & \textbf{-}  \\
		Proposed    & \textbf{0.9704} 	&\textbf{0.9409} 	&\textbf{0.8754 } 	&\textbf{0.9180} \\
		
		\bottomrule[.15em]
	\end{tabular}

\end{table}

\begin{table*}

	\centering
	\renewcommand{\arraystretch}{1.25}
	\caption{The SRCC results of training ResNet50 with and without the staircase structure on individual database or using the iterative mixed database training strategy. The best performing model is highlighted in each column. SS means the staircase structure.}
	\label{ablation_result}
	\begin{tabular}{ccccc|cccccc}
		\toprule[.15em]
		Backbone&IMDT& SS& Para. (M) & GFlops & CLIVE & BID & KonIQ10k & SPQA & FLIVE & FLIVE Patch  \\
		
		\hline
		
		ResNet50 &$\surd$&$\surd$ & 31.799 &  10.382  &\textbf{0.8992} 	 	&\textbf{0.9128} 	 	&\textbf{0.9209} 	 	&\textbf{0.9238} 	 &	\textbf{0.5821}  	& \textbf{0.7679} 	 \\
		
		ResNet50 &$\times$&$\surd$ &30.487& 10.381 & 0.8624	 & 0.8724  & 0.9186 & 0.9208 & 0.5733  & 0.7669   \\
		
		ResNet50 &$\surd$&$\times$  & 25.082 &  8.388 & 0.8981  & 0.9102  &0.9189  & 0.9237   & 0.5802 &0.7643 	 \\
		ResNet50 &$\times$&$\times$  & 23.770 &  8.387 &0.8479  & 0.8598 &0.9100  &0.9163  &0.5637 	& 0.7646  \\
		ResNet101 &$\times$&$\times$   & 42.763 & 15.984 &0.8546  &0.8653  &0.9155  &0.9186   &0.5654  &0.7653   \\
		
		\bottomrule[.15em]
	\end{tabular}
\end{table*}

\subsubsection{Visualization of the Evaluation Effect of the Proposed Model}

\begin{figure}[!t]
	\centering
	\includegraphics[height=2in]{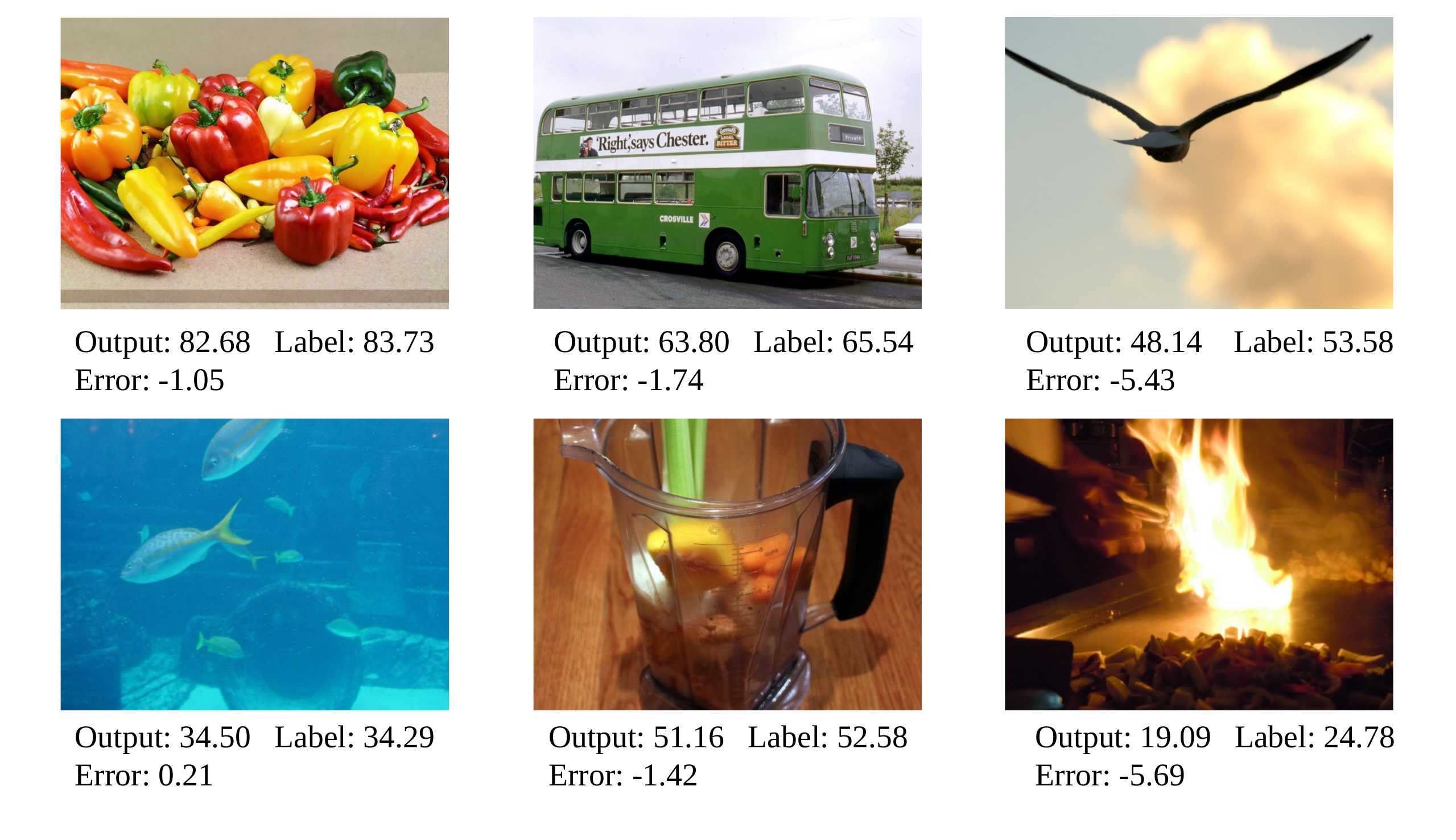}
	\caption{The comparison between the quality scores evaluated by proposed model and the ground truth on some example images in the KonIQ10k \cite{hosu2020koniq} database. ‘Output’ represents the quality score evaluated by the proposed model, ‘Label’ represents the ground-truth quality score obtained from the subjective experiments, and ‘Error’ represents the difference between the predicted quality score and the ground-truth quality score. Some failing examples are shown in the last column.}
	\label{example_images}
\end{figure}

We show some example images with their quality scores predicted by the proposed model in Fig. \ref{example_images}. From Fig. \ref{example_images}, we can observe that the proposed model can effectively predict the quality scores of in-the-wild images, and the predicted scores are mostly consistent with the MOSs assessed by the subjects. Some failing examples are shown in the last column in Fig. \ref{example_images}. It is seen that the proposed model tends to give lower scores to the images with low visibility and blurred content. However, low visibility and blurred content may include artistic images, which can also give users a high-quality experience. Hence, the proposed model also needs to enhance the understanding of the image aesthetics, which is very important but challenging for the quality assessment of in-the-wild images.

\subsection{Ablation Studies}
\label{ablation_studies}
The proposed model introduces a staircase structure to integrate features from intermediate layers and utilizes the IMDT strategy to increase training samples and the diversity of image content and distortion. In order to investigate the effectiveness of each module, we conduct two experiments. First, we train the ResNet50 with the staircase structure on individual databases and the ResNet50 using the IMDT strategy respectively. We also train ResNet50 and ResNet101 on individual databases as the baseline. Second, since the IMDT strategy takes advantage of the information of other databases, it is necessary to compare it with the methods that can also utilize other databases like transfer learning. Hence, we train ResNet50 with the staircase structure on the small-scale databases (i.e. CLVIE and BID) by initializing the weights pretrained on a large-scale IQA database (i.e. KonIQ10k, SPAQ, etc.) to further validate the effectiveness of the IMDT strategy. The results of two experiments are shown in Table \ref{ablation_result} and Fig. \ref{transfer_learning} respectively.

\subsubsection{The Effects of the Staircase Structure and IMDT} From Table \ref{ablation_result}, it is observed that the performances of the model without the staircase structure and the model without using the IMDT strategy are both inferior to the complete model and superior to the baseline model, which indicates that both the staircase structure and the IMDT strategy make contributions to the overall model.
Interestingly, we notice that the two modules have different contributions to different databases. For CLVIE and BID databases, the IMDT strategy can greatly improve the model performance. The reason is that the number of images in CLVIE and BID databases is far smaller than other databases and the IMDT strategy allows the model to learn useful feature representation from other databases, where the image content and distortions are more abundant and diverse. 
For large-scale databases like KonIQ10k, SPAQ, etc., we find the model with the staircase structure achieves competitive performance with the model trained by the IMDT strategy. The reason is that the image content and distortions in these databases are diverse enough to train a deep model. Then, the staircase structure can help the model incorporate features from intermediate layers, which allows the model to learn better feature representation for quality evaluation too. 
Then, the performance of ResNet50 with the staircase structure is better than ResNet101 on all six databases, and the number of parameters of ResNet101 is larger than the former, which indicates that performance improvement contributed by the staircase structure is due to better feature representation rather than the added parameters. 

\subsubsection{The IMDT vs. Transfer Learning Method} 
\begin{figure}[!t]
	\centering
	\includegraphics[height=2in]{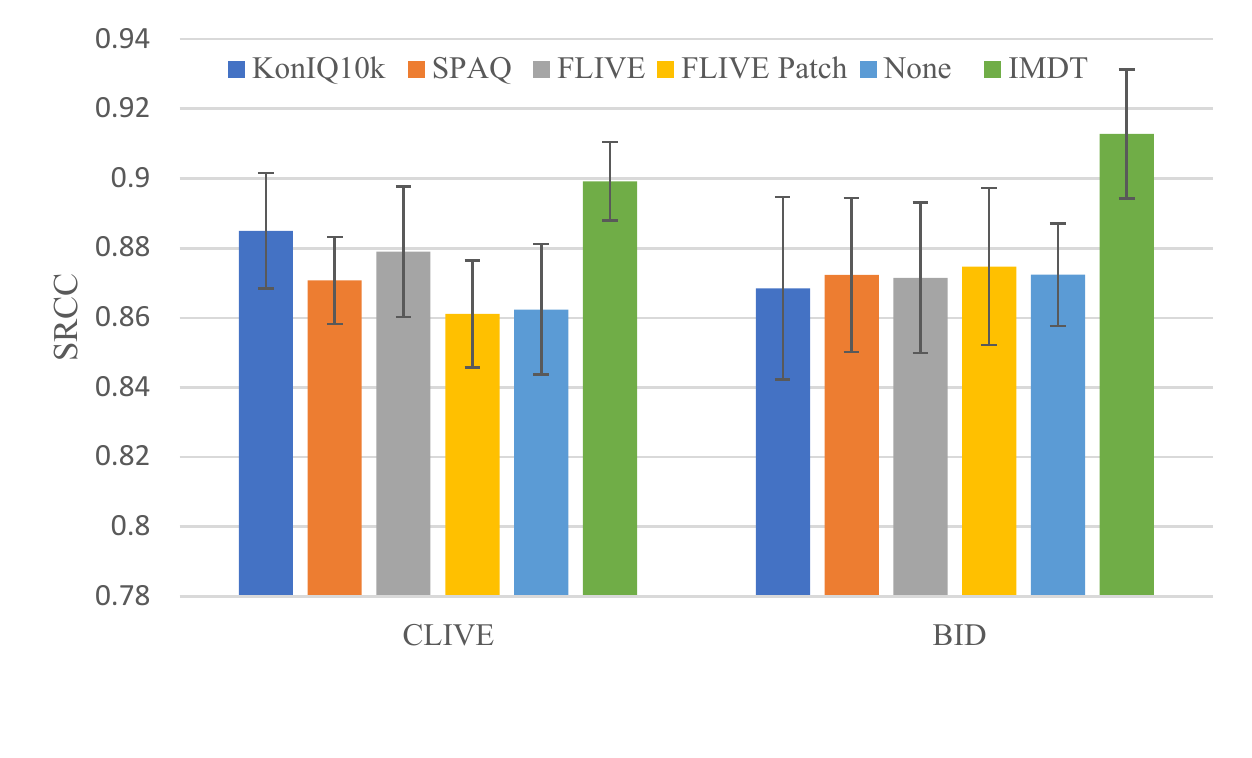}
	\caption{SRCC of ResNet50 with the staircase structure on the CLIVE and BID databases by initializing the weights pretrained on the other four databases. We also list the performance of the model without using the pretrained model and trained by the IMDT strategy for comparison.}
	\label{transfer_learning}
\end{figure}

From Fig. \ref{transfer_learning}, we observe that the models pretrained on other large-scale IQA databases do improve the model performance, but no matter which database is pretrained on, their performance is far lower than the model trained by the IMDT strategy. There are two advantages when adopting the IMDT strategy rather than using the model weights pretrained on other databases. First, the IMDT strategy allows the model to be trained on multiple databases simultaneously, while the pretrained model is only trained on one database. Hence, the IMDT strategy can take full advantage of more diverse image content and distortions from multiple databases. Second, the number of training images is not increased when using the transfer learning method, and there is still a risk that the deep model may overfit on the small database during the training stage. So, the IMDT strategy can further promote model performance than the transfer learning method.

\subsection{Cross-Database Evaluation}
\label{cross_dataset_evaluation}

\label{cross_dataset_evaluation}
\begin{table*}
	
	\centering
	\renewcommand{\arraystretch}{1.25}
	\caption{The SRCC results of the cross-database evaluation. We evaluate the proposed model by training it on five databases simultaneously and then test it on the left database. Note that the trained models include five image quality regressors, which correspond to the five training databases respectively. Therefore, we test these five image quality regressors on the left database. The database name in each column indicates the database which the regressor is trained on. The values in the bracket represent the performance gain when compared with the ResNet50 with staircase structure trained on the corresponding column database name and tested on the corresponding row database name. ``Ensemble'' means the quality scores are avegrated results of five regressors.}
	\label{result_cross_database_evaluation}
	\resizebox{1\textwidth}{!}{
	\begin{tabular}{c|ccccccc}
		\toprule[.15em]
		 \multirow{2}{*}{Test database} & \multicolumn{6}{c}{Training database}  \\
		
		 & CLIVE&BID&KonIQ10k&SPQA&FLIVE&FLIVE Patch &Ensemble   \\
		\hline
		
		CLIVE   &- & 0.8603   ({\color{blue}+0.0669})  &	0.8450  ({\color{blue}+0.0497})	&0.8369	({\color{blue}+0.0364})	&0.8510	({\color{blue}+0.0876}) 	& 0.8007	 ({\color{blue}+0.0645})	& \textbf{0.8663} \\
		BID  &\textbf{0.8882}	({\color{blue}+0.0029}) & -		& 0.8767	 ({\color{blue}+0.0634})	&0.8434	 ({\color{blue}+0.0353})&0.8488	({\color{blue}+0.0881}) 	&0.7969 	({\color{blue}+0.0717})	&	0.8730  \\
		
		KonIQ10k   & 0.8223	 ({\color{blue}+0.0550}) 	& 0.8281 	 ({\color{blue}+0.0939})  	& -		& 0.8434	({\color{blue}+0.0282})	& 0.8266  	 ({\color{blue}+0.0445}) & 0.7692	({\color{blue}+0.0437})	&\textbf{0.8483} \\			
		SPQA& 0.8843   ({\color{blue}+0.0164})	& 0.8781 	( {\color{blue}+0.0455})	& 0.8690	 ({\color{blue}+0.0055})	& -		& 0.8826	({\color{blue}+0.0289})  & 0.8868  ({\color{blue}+0.0281}) &	\textbf{0.8914}\\
		FLIVE & 0.4710   ({\color{blue}+0.0857})	&  0.4600   ({\color{blue}+0.1179})	& 0.4735	  ({\color{blue}+0.0055})   & 0.4976 	 ({\color{blue}+0.0544}) 	& - 	& 0.4874 	({\color{blue}+0.0023}) &	\textbf{0.5061}\\
		FLIVE patch   & 0.6153	 ({\color{blue}+0.0409})   & 0.5816 	 ({\color{blue}+0.0571}) 	& 0.5442 	 ({\color{blue}+0.1103}) 	& 0.6362   ({\color{blue}+0.0219})  & \textbf{0.6537}    ({\color{blue}+0.0059}) & - 	&	0.6319  \\
		
		\bottomrule[.15em]
	\end{tabular}
 }
\end{table*}

As discussed above, the quality of in-the-wild images depends on their content and distortions. However, due to the effects of different cameras, shooting environments, photographic skills, etc., the image content and distortions may vary greatly in practical applications. For the given in-the-wild IQA database, it only includes limited types of image content and distortions. Moreover, the images in such IQA databases may follow some specific rules, for example, they were taken by a limited type of cameras \cite{ciancio2011no}\cite{fang2020perceptual} or they were selected from the Internet according to the distribution rule of certain image attributes \cite{hosu2020koniq}. As a result, there is a risk that the models trained on these databases may learn the feature representation only suitable for specific databases. Therefore, cross-database evaluation is very important for IQA models since it can reflect their generalization ability of models for images obtained from totally different manners. In this section, we test the generalization ability of the proposed model via cross-database evaluation.

Since the proposed model can be trained on multiple databases, we adopt the leave-one-database-out method to test the performance of cross-database evaluation. Specifically, we evaluate the proposed model by training it on five databases simultaneously and then test it on the left database \textbf{without any fine-tuning}. Since the trained models include five image quality regressors, we test all of them and the ensemble of them on the left database. In addition, we also train the ResNet50 with staircase structure on each database (without using the IMDT strategy) and then test it on the other five databases respectively to show the effect of the IMDT strategy on cross-database evaluation. We list the results of the cross-database evaluation in Table \ref{result_cross_database_evaluation}. 

From Table \ref{result_cross_database_evaluation}, we first observe that the proposed model achieves an excellent performance of cross-database evaluation. Specifically, besides using the image quality regressors trained on the FLIVE patch database, the SRCC values tested on the CLIVE, BID, KonIQ10k, and SPAQ databases all exceed 0.8, which have surpassed most BIQA models trained on the corresponding databases as shown in Table \ref{performance}. It demonstrates that the proposed model does learn a strong feature representation for quality evaluation, and this kind of feature representation is robust to images sampled from various distributions. For the latest IQA model HyperIQA \cite{su2020blindly}, when it is trained on the KonIQ10k database and then tested on the CLIVE and BID databases, its SRCC values are 0.785 and 0.819 respectively, which are far lower than our model (0.846 and 0.868).
Second, it is seen that the IMDT strategy greatly improves the performance of cross-database evaluation when compared with the model without using the IMDT strategy. Note that one motivation of the IMDT strategy is to make the feature extraction module learn a more general feature representation from multiple databases, and the results of cross-database evaluation also prove that the IMDT strategy does have this function. Third, the performance of the ensemble result is normally better than the results outputted by the regressor trained by a specific database, which mainly because the ensemble manner can make the regressors take advantage of all the training samples and is more robust to unseen images.

It is noticed that the performance of using the model trained on the FLIVE patch database is lower than that of the models trained on other databases. The reason is that the resolutions of image patches in the FLIVE patch database are relatively small, and most of them do not include complete content with semantic information. Therefore, the perceptual quality scores rated by users are mainly according to the distortions rather than content information. Though the feature extraction module trained by the IMDT strategy can extract the feature representation of both image content and distortions, the image quality regressor trained on the FLIVE patch database will focus on the features related to image distortions and ignore the features related to image content. But for the other four databases, the image content is an important factor for their perceptual quality, so the image quality regressors trained by them are effective for both image distortion and content.
\vspace{-0.3cm}

\subsection{The Effects of Different Backbones}
In this section, we test different backbones to show the effect of the backbone networks on the performance of the model. Specifically, we train three CNN models, MobileNetV2 \cite{sandler2018mobilenetv2}, ResNext50 \cite{xie2017aggregated}, and ResNext101 \cite{xie2017aggregated} with the staircase structure on the KonIQ10k database. MobileNetV2 is a lightweight CNN model for mobile applications while ResNext50 and ResNext101 are two more powerful CNN structures than ResNet50. The results are listed in Table \ref{performance_backbones}. From Table \ref{performance_backbones}, we observe that the performance of ResNext50 is superior to the ResNet50 though they have a similar number of parameters. Comparing the performances of ResNext50 and ResNext101, we find that as the number of network layers increases, the performance of the proposed model is also significantly improved. MobileNetV2 has ten times fewer parameters than ResNet50, but the SRCC value of MobileNetV2 is only 0.0123 less than ResNet50. Therefore, the staircase structure is a flexible and effective module for BIQA, which can be integrated with popular CNN models, and we can choose the corresponding model to meet requirements such as a greater emphasis on high accuracy or faster running time.
\begin{table}
	\footnotesize
	\centering
	\renewcommand{\arraystretch}{1.25}
	\caption{The performance of different backbones with the staircase structure on the KonIQ10k database.}
	\label{performance_backbones}
	\begin{tabular}{c|cccc}
		\toprule[.15em]	
		
		Backbones &  MobileNetV2  & ResNet50& ResNext50 & ResNext101   \\
		\hline

		SRCC    &0.9063&0.9186&0.9202&\textbf{0.9304} \\
		PLCC  &0.9235 &0.9346&0.9363&\textbf{0.9435}  \\
			
		\bottomrule[.15em]
		
	\end{tabular}
\vspace{-0.3cm}
\end{table}

\subsection{The Effects of Features Extracted from Different Stages}
As illustrated in Fig. \ref{framework}, the staircase structure consists of four convolutional paths to fuse the features extracted from different stages to the final feature representation. So, it is necessary to investigate the effect of these features on the final performance. In this section, we train the backbone network (i.e. ResNet50) with these four convolutional paths individually on the KonIQ10k database to verify the contributions of features extracted from different stages to the overall quality evaluation. We list the results in Table \ref{performance_backbones}. First, when comparing the performances of $Path4$ to $Path1$, we find the performance increases monotonously as the features extracted from Stage4 to Stage1 are added in sequence to the model, which indicates that the features extracted from all stages make contributions to the overall performance. Then, we observe that the performance gains of $Path3$ to $Path4$ and $Path4$ to fusing no features are larger than the performance gains of $Path1$ to $Path2$ and $Path2$ to $Path3$, which means the features extracted from Stage3 and Stage4 are more important to the image quality evaluation. Finally, the performances of $Path4$ to $Path1$ are all inferior to the proposed staircase structure, which indicates that combing the fused features from these convolutional paths can further improve the performance of the model.

\begin{table}
	\footnotesize
	\centering
	\renewcommand{\arraystretch}{1.25}
	\caption{The performance of ResNet50 with different convolutional paths on the KonIQ10k database. S in the first row means Stage.}
	\label{performance_backbones}
		\begin{tabular}{c|cccc|cc}
			\toprule[.15em]	
			
			 Convolutional Path & S1  & S2 & S3 & S4 & SRCC & PLCC   \\
			\hline
			
			$Path$ 1    &$\surd$ &$\surd$  & $\surd$ &$\surd$ & 0.9169  & 0.9324 \\
			$Path$ 2    & $\times$&$\surd$ &$\surd$  &$\surd$ & 0.9157 &  0.9328 \\
			$Path$ 3    & $\times$& $\times$ & $\surd$ &$\surd$ & 0.9146 & 0.9316  \\
			$Path$ 4    &$\times$ & $\times$ & $\times$ & $\surd$&0.9128  & 0.9289  \\
			None  &$\times$ & $\times$ & $\times$ & $\times$&0.9100  & 0.9259  \\
			\bottomrule[.15em]
			
		\end{tabular}
\end{table}

\subsection{The Choose of Regressor}
Since the proposed IMDT adds multiple regressors to train the IQA model on databases, it is necessary to choose a proper regressor when used in real-world scenarios. In practice, there are two situations that may affect how to choose the regressor. The first one is that we have already built an IQA database to solve a specific IQA problem (i.e. quality assessment for social media images). Then, we can train the built IQA database with other existing IQA databases (i.e. the six IQA databases used in the paper) and we can use the regressor of the built IQA databases since this regressor is most closed to the situation we need to solve. The second one is that we do not have any private IQA databases and we only use the public IQA databases to train the IQA models. In this situation, we recommend using the ensemble results of the outputs of six regressors, since it can make the regressor take advantage of all the training samples and are more robust to unseen images, which is demonstrated in Section \ref{cross_dataset_evaluation}.

\subsection{Discussion}
BIQA is a very challenging problem because of the high-dimensional space of the image content and distortion as well as the lack of reference images. As a result, a mass of labeled data is necessary to train an excellent BIQA model by supervised learning. However, subject IQA experiments are extremely time-consuming and cumbersome, which causes it difficult and even impossible to construct an ideal IQA database that includes enough images and all kinds of distortion types. In this paper, on one hand, we propose a new feature extraction module (i.e. staircase structure) for BIQA, which hierarchically fuses the features from intermediate layers into the final feature representation. So, the model can easily utilize the low-level visual information in the early stage of the CNN model, which is very useful for image distortion modeling but is rarely contained in the features from the final stage of the CNN model. With the help of the proposed staircase structure, the model can learn more effective quality-aware features even with a small-scale IQA database. On the other hand, we propose a new training method to train the model on multiple existing IQA databases and make the model learn a more general feature representation ability. Hence, when we need to propose a new kind of IQA method, we may take advantage of other IQA databases through the proposed IMDT strategy and decrease the cost of constructing subjective IQA databases. Moreover, the IMDT strategy is helpful to recent emerging studies in IQA such as meta-learning \cite{zhu2020metaiqa} \cite{zhu2021generalizable}, continual learning \cite{zhang2022continual,zhang2021task} etc., which also try to utilize the model trained on existing databases to a new database and decrease the cost of training on the new database.

\section{Conclusion}
\label{conclusion}
In this paper, we propose a new BIQA model for in-the-wild images. The proposed model consists of two novel modules: the staircase structure and the iterative mixed database training (IMDT) strategy. The staircase structure makes the model integrate the features from intermediate layers into the final feature representation, so the model can make full use of visual information from low-level to high-level. The IMDT strategy allows the model to be trained on multiple databases simultaneously, which makes the model benefit from the increase in both training images and image content and distortion diversity and learn a more general feature representation. Experimental results show that the proposed model outperforms other state-of-art BIQA models on six in-the-wild IQA databases, and also achieves an excellent performance in the cross-database evaluation, which demonstrate the effectiveness and generalizability of the proposed model.

\bibliographystyle{IEEEtran}
\bibliography{bare_jrnl}

\begin{thebibliography}{10}
\providecommand{\url}[1]{#1}
\csname url@samestyle\endcsname
\providecommand{\newblock}{\relax}
\providecommand{\bibinfo}[2]{#2}
\providecommand{\BIBentrySTDinterwordspacing}{\spaceskip=0pt\relax}
\providecommand{\BIBentryALTinterwordstretchfactor}{4}
\providecommand{\BIBentryALTinterwordspacing}{\spaceskip=\fontdimen2\font plus
\BIBentryALTinterwordstretchfactor\fontdimen3\font minus
  \fontdimen4\font\relax}
\providecommand{\BIBforeignlanguage}[2]{{%
\expandafter\ifx\csname l@#1\endcsname\relax
\typeout{** WARNING: IEEEtran.bst: No hyphenation pattern has been}%
\typeout{** loaded for the language `#1'. Using the pattern for}%
\typeout{** the default language instead.}%
\else
\language=\csname l@#1\endcsname
\fi
#2}}
\providecommand{\BIBdecl}{\relax}
\BIBdecl

\bibitem{sun2022blind}
W.~Sun, H.~Duan, X.~Min, L.~Chen, and G.~Zhai, ``Blind quality assessment for
  in-the-wild images via hierarchical feature fusion strategy,'' in \emph{2022
  IEEE International Symposium on Broadband Multimedia Systems and Broadcasting
  (BMSB)}.\hskip 1em plus 0.5em minus 0.4em\relax IEEE, 2022, pp. 01--06.

\bibitem{zhang2020data}
X.~Zhang, S.~Kwong, and C.-C.~J. Kuo, ``Data-driven transform-based compressed
  image quality assessment,'' \emph{IEEE Transactions on Circuits and Systems
  for Video Technology}, vol.~31, no.~9, pp. 3352--3365, 2020.

\bibitem{wu2020subjective}
Q.~Wu, L.~Wang, K.~N. Ngan, H.~Li, F.~Meng, and L.~Xu, ``Subjective and
  objective de-raining quality assessment towards authentic rain image,''
  \emph{IEEE Transactions on Circuits and Systems for Video Technology},
  vol.~30, no.~11, pp. 3883--3897, 2020.

\bibitem{zhai2021perceptual}
G.~Zhai, W.~Sun, X.~Min, and J.~Zhou, ``Perceptual quality assessment of
  low-light image enhancement,'' \emph{ACM Transactions on Multimedia
  Computing, Communications, and Applications (TOMM)}, vol.~17, no.~4, pp.
  1--24, 2021.

\bibitem{zhou2021omnidirectional}
Y.~Zhou, Y.~Sun, L.~Li, K.~Gu, and Y.~Fang, ``Omnidirectional image quality
  assessment by distortion discrimination assisted multi-stream network,''
  \emph{IEEE Transactions on Circuits and Systems for Video Technology}, 2021.

\bibitem{moorthy2011blind}
A.~K. Moorthy and A.~C. Bovik, ``Blind image quality assessment: From natural
  scene statistics to perceptual quality,'' \emph{IEEE transactions on Image
  Processing}, vol.~20, no.~12, pp. 3350--3364, 2011.

\bibitem{mittal2012no}
A.~Mittal, A.~K. Moorthy, and A.~C. Bovik, ``No-reference image quality
  assessment in the spatial domain,'' \emph{IEEE Transactions on image
  processing}, vol.~21, no.~12, pp. 4695--4708, 2012.

\bibitem{gu2014using}
K.~Gu, G.~Zhai, X.~Yang, and W.~Zhang, ``Using free energy principle for blind
  image quality assessment,'' \emph{IEEE Transactions on Multimedia}, vol.~17,
  no.~1, pp. 50--63, 2014.

\bibitem{min2017blind}
X.~Min, K.~Gu, G.~Zhai, J.~Liu, X.~Yang, and C.~W. Chen, ``Blind quality
  assessment based on pseudo-reference image,'' \emph{IEEE Transactions on
  Multimedia}, vol.~20, no.~8, pp. 2049--2062, 2017.

\bibitem{min2018blind}
X.~{Min}, G.~{Zhai}, K.~{Gu}, Y.~{Liu}, and X.~{Yang}, ``Blind image quality
  estimation via distortion aggravation,'' \emph{IEEE Transactions on
  Broadcasting}, vol.~64, no.~2, pp. 508--517, 2018.

\bibitem{zhai2019free}
G.~Zhai, X.~Min, and N.~Liu, ``Free-energy principle inspired visual quality
  assessment: An overview,'' \emph{Digital Signal Processing}, vol.~91, pp.
  11--20, 2019.

\bibitem{saad2012blind}
M.~A. Saad, A.~C. Bovik, and C.~Charrier, ``Blind image quality assessment: A
  natural scene statistics approach in the dct domain,'' \emph{IEEE
  transactions on Image Processing}, vol.~21, no.~8, pp. 3339--3352, 2012.

\bibitem{wu2022disentangling}
H.~Wu, L.~Liao, C.~Chen, J.~Hou, A.~Wang, W.~Sun, Q.~Yan, and W.~Lin,
  ``Disentangling aesthetic and technical effects for video quality assessment
  of user generated content,'' \emph{arXiv preprint arXiv:2211.04894}, 2022.

\bibitem{simonyan2015very}
K.~{Simonyan} and A.~{Zisserman}, ``Very deep convolutional networks for
  large-scale image recognition,'' in \emph{ICLR 2015 : International
  Conference on Learning Representations 2015}, 2015.

\bibitem{he2016deep}
K.~{He}, X.~{Zhang}, S.~{Ren}, and J.~{Sun}, ``Deep residual learning for image
  recognition,'' in \emph{2016 IEEE Conference on Computer Vision and Pattern
  Recognition (CVPR)}, 2016, pp. 770--778.

\bibitem{kang2014convolutional}
L.~Kang, P.~Ye, Y.~Li, and D.~Doermann, ``Convolutional neural networks for
  no-reference image quality assessment,'' in \emph{Proceedings of the IEEE
  conference on computer vision and pattern recognition}, 2014, pp. 1733--1740.

\bibitem{ying2020from}
Z.~{Ying}, H.~{Niu}, P.~{Gupta}, D.~{Mahajan}, D.~{Ghadiyaram}, and A.~{Bovik},
  ``From patches to pictures (paq-2-piq): Mapping the perceptual space of
  picture quality,'' in \emph{2020 IEEE/CVF Conference on Computer Vision and
  Pattern Recognition (CVPR)}, 2020, pp. 3575--3585.

\bibitem{zeiler2014visualizing}
M.~D. Zeiler and R.~Fergus, ``Visualizing and understanding convolutional
  networks,'' in \emph{European conference on computer vision}.\hskip 1em plus
  0.5em minus 0.4em\relax Springer, 2014, pp. 818--833.

\bibitem{ranjan2017hyperface}
R.~Ranjan, V.~M. Patel, and R.~Chellappa, ``Hyperface: A deep multi-task
  learning framework for face detection, landmark localization, pose
  estimation, and gender recognition,'' \emph{IEEE Transactions on Pattern
  Analysis and Machine Intelligence}, vol.~41, no.~1, pp. 121--135, 2017.

\bibitem{liu2019unsupervised}
Y.~Liu, K.~Gu, Y.~Zhang, X.~Li, G.~Zhai, D.~Zhao, and W.~Gao, ``Unsupervised
  blind image quality evaluation via statistical measurements of structure,
  naturalness, and perception,'' \emph{IEEE Transactions on Circuits and
  Systems for Video Technology}, vol.~30, no.~4, pp. 929--943, 2019.

\bibitem{xu2016blind}
J.~Xu, P.~Ye, Q.~Li, H.~Du, Y.~Liu, and D.~Doermann, ``Blind image quality
  assessment based on high order statistics aggregation,'' \emph{IEEE
  Transactions on Image Processing}, vol.~25, no.~9, pp. 4444--4457, 2016.

\bibitem{xu2016multi}
L.~Xu, J.~Li, W.~Lin, Y.~Zhang, L.~Ma, Y.~Fang, and Y.~Yan, ``Multi-task rank
  learning for image quality assessment,'' \emph{IEEE Transactions on Circuits
  and Systems for Video Technology}, vol.~27, no.~9, pp. 1833--1843, 2016.

\bibitem{bosse2017deep}
S.~Bosse, D.~Maniry, K.-R. M{\"u}ller, T.~Wiegand, and W.~Samek, ``Deep neural
  networks for no-reference and full-reference image quality assessment,''
  \emph{IEEE Transactions on Image Processing}, vol.~27, no.~1, pp. 206--219,
  2017.

\bibitem{ma2017end}
K.~Ma, W.~Liu, K.~Zhang, Z.~Duanmu, Z.~Wang, and W.~Zuo, ``End-to-end blind
  image quality assessment using deep neural networks,'' \emph{IEEE
  Transactions on Image Processing}, vol.~27, no.~3, pp. 1202--1213, 2017.

\bibitem{zhang2020blind}
W.~{Zhang}, K.~{Ma}, J.~{Yan}, D.~{Deng}, and Z.~{Wang}, ``Blind image quality
  assessment using a deep bilinear convolutional neural network,'' \emph{IEEE
  Transactions on Circuits and Systems for Video Technology}, vol.~30, no.~1,
  pp. 36--47, 2020.

\bibitem{su2020blindly}
S.~Su, Q.~Yan, Y.~Zhu, C.~Zhang, X.~Ge, J.~Sun, and Y.~Zhang, ``Blindly assess
  image quality in the wild guided by a self-adaptive hyper network,'' in
  \emph{Proceedings of the IEEE/CVF Conference on Computer Vision and Pattern
  Recognition}, 2020, pp. 3667--3676.

\bibitem{zhu2021generalizable}
H.~Zhu, L.~Li, J.~Wu, W.~Dong, and G.~Shi, ``Generalizable no-reference image
  quality assessment via deep meta-learning,'' \emph{IEEE Transactions on
  Circuits and Systems for Video Technology}, 2021.

\bibitem{ke2021musiq}
J.~Ke, Q.~Wang, Y.~Wang, P.~Milanfar, and F.~Yang, ``Musiq: Multi-scale image
  quality transformer,'' in \emph{Proceedings of the IEEE/CVF International
  Conference on Computer Vision}, 2021, pp. 5148--5157.

\bibitem{ye2012unsupervised}
P.~Ye, J.~Kumar, L.~Kang, and D.~Doermann, ``Unsupervised feature learning
  framework for no-reference image quality assessment,'' in \emph{2012 IEEE
  conference on computer vision and pattern recognition}.\hskip 1em plus 0.5em
  minus 0.4em\relax IEEE, 2012, pp. 1098--1105.

\bibitem{zhang2014training}
L.~Zhang, Z.~Gu, X.~Liu, H.~Li, and J.~Lu, ``Training quality-aware filters for
  no-reference image quality assessment,'' \emph{IEEE MultiMedia}, vol.~21,
  no.~4, pp. 67--75, 2014.

\bibitem{liu2017rankiqa}
X.~{Liu}, J.~van~de {Weijer}, and A.~D. {Bagdanov}, ``Rankiqa: Learning from
  rankings for no-reference image quality assessment,'' in \emph{2017 IEEE
  International Conference on Computer Vision (ICCV)}, 2017, pp. 1040--1049.

\bibitem{ma2017dipiq}
K.~{Ma}, W.~{Liu}, T.~{Liu}, Z.~{Wang}, and D.~{Tao}, ``dipiq: Blind image
  quality assessment by learning-to-rank discriminable image pairs,''
  \emph{IEEE Transactions on Image Processing}, vol.~26, no.~8, pp. 3951--3964,
  2017.

\bibitem{gu2017learning}
K.~Gu, D.~Tao, J.-F. Qiao, and W.~Lin, ``Learning a no-reference quality
  assessment model of enhanced images with big data,'' \emph{IEEE transactions
  on neural networks and learning systems}, vol.~29, no.~4, pp. 1301--1313,
  2017.

\bibitem{ye2014beyond}
P.~Ye, J.~Kumar, and D.~Doermann, ``Beyond human opinion scores: Blind image
  quality assessment based on synthetic scores,'' in \emph{Proceedings of the
  IEEE Conference on Computer Vision and Pattern Recognition}, 2014, pp.
  4241--4248.

\bibitem{po2019novel}
L.-M. Po, M.~Liu, W.~Y. Yuen, Y.~Li, X.~Xu, C.~Zhou, P.~H. Wong, K.~W. Lau, and
  H.-T. Luk, ``A novel patch variance biased convolutional neural network for
  no-reference image quality assessment,'' \emph{IEEE Transactions on Circuits
  and Systems for Video Technology}, vol.~29, no.~4, pp. 1223--1229, 2019.

\bibitem{krasula2019training}
L.~Krasula, Y.~Baveye, and P.~Le~Callet, ``Training objective image and video
  quality estimators using multiple databases,'' \emph{IEEE Transactions on
  Multimedia}, vol.~22, no.~4, pp. 961--969, 2019.

\bibitem{li2021unified}
D.~Li, T.~Jiang, and M.~Jiang, ``Unified quality assessment of in-the-wild
  videos with mixed datasets training,'' \emph{International Journal of
  Computer Vision}, vol. 129, no.~4, pp. 1238--1257, 2021.

\bibitem{zhang2021uncertainty}
W.~Zhang, K.~Ma, G.~Zhai, and X.~Yang, ``Uncertainty-aware blind image quality
  assessment in the laboratory and wild,'' \emph{IEEE Transactions on Image
  Processing}, vol.~30, pp. 3474--3486, 2021.

\bibitem{szegedy2015going}
C.~{Szegedy}, W.~{Liu}, Y.~{Jia}, P.~{Sermanet}, S.~{Reed}, D.~{Anguelov},
  D.~{Erhan}, V.~{Vanhoucke}, and A.~{Rabinovich}, ``Going deeper with
  convolutions,'' in \emph{2015 IEEE Conference on Computer Vision and Pattern
  Recognition (CVPR)}, 2015, pp. 1--9.

\bibitem{ren2015faster}
S.~Ren, K.~He, R.~Girshick, and J.~Sun, ``Faster r-cnn: Towards real-time
  object detection with region proposal networks,'' in \emph{Advances in neural
  information processing systems}, 2015, pp. 91--99.

\bibitem{gao2017deepsim}
F.~Gao, Y.~Wang, P.~Li, M.~Tan, J.~Yu, and Y.~Zhu, ``Deepsim: Deep similarity
  for image quality assessment,'' \emph{Neurocomputing}, vol. 257, pp.
  104--114, 2017.

\bibitem{ali2017image}
S.~Ali~Amirshahi, M.~Pedersen, and S.~X. Yu, ``Image quality assessment by
  comparing cnn features between images,'' \emph{Electronic Imaging}, vol.
  2017, no.~12, pp. 42--51, 2017.

\bibitem{kingma2015adam}
D.~P. {Kingma} and J.~L. {Ba}, ``Adam: A method for stochastic optimization,''
  in \emph{ICLR 2015 : International Conference on Learning Representations
  2015}, 2015.

\bibitem{ghadiyaram2016massive}
D.~{Ghadiyaram} and A.~C. {Bovik}, ``Massive online crowdsourced study of
  subjective and objective picture quality,'' \emph{IEEE Transactions on Image
  Processing}, vol.~25, no.~1, pp. 372--387, 2016.

\bibitem{ciancio2011no}
A.~{Ciancio}, A.~L. N.~T. da~{Costa}, E.~A.~B. da~{Silva}, A.~{Said},
  R.~{Samadani}, and P.~{Obrador}, ``No-reference blur assessment of digital
  pictures based on multifeature classifiers,'' \emph{IEEE Transactions on
  Image Processing}, vol.~20, no.~1, pp. 64--75, 2011.

\bibitem{hosu2020koniq}
V.~{Hosu}, H.~{Lin}, T.~{Sziranyi}, and D.~{Saupe}, ``Koniq-10k: An
  ecologically valid database for deep learning of blind image quality
  assessment,'' \emph{IEEE Transactions on Image Processing}, vol.~29, pp.
  4041--4056, 2020.

\bibitem{fang2020perceptual}
Y.~{Fang}, H.~{Zhu}, Y.~{Zeng}, K.~{Ma}, and Z.~{Wang}, ``Perceptual quality
  assessment of smartphone photography,'' in \emph{2020 IEEE/CVF Conference on
  Computer Vision and Pattern Recognition (CVPR)}, 2020, pp. 3677--3686.

\bibitem{murray2012ava}
N.~{Murray}, L.~{Marchesotti}, and F.~{Perronnin}, ``Ava: A large-scale
  database for aesthetic visual analysis,'' in \emph{2012 IEEE Conference on
  Computer Vision and Pattern Recognition}, 2012, pp. 2408--2415.

\bibitem{everingham2010the}
M.~{Everingham}, L.~{Gool}, C.~K. {Williams}, J.~{Winn}, and A.~{Zisserman},
  ``The pascal visual object classes (voc) challenge,'' \emph{International
  Journal of Computer Vision}, vol.~88, no.~2, pp. 303--338, 2010.

\bibitem{kosti2017emotic}
R.~{Kosti}, J.~M. {Alvarez}, A.~{Recasens}, and A.~{Lapedriza}, ``Emotic:
  Emotions in context dataset,'' in \emph{2017 IEEE Conference on Computer
  Vision and Pattern Recognition Workshops (CVPRW)}, 2017, pp. 2309--2317.

\bibitem{mavridaki2014no}
E.~{Mavridaki} and V.~{Mezaris}, ``No-reference blur assessment in natural
  images using fourier transform and spatial pyramids,'' in \emph{2014 IEEE
  International Conference on Image Processing (ICIP)}, 2014, pp. 566--570.

\bibitem{sheikh2005live}
H.~R. Sheikh, Z.~Wang, L.~Cormack, and A.~C. Bovik, ``Live image quality
  assessment database release 2 (2005),'' \emph{URL http://live. ece. utexas.
  edu/research/quality}, 2005.

\bibitem{larson2010most}
E.~C. Larson and D.~M. Chandler, ``Most apparent distortion: full-reference
  image quality assessment and the role of strategy,'' \emph{Journal of
  electronic imaging}, vol.~19, no.~1, p. 011006, 2010.

\bibitem{lin2019kadid}
H.~Lin, V.~Hosu, and D.~Saupe, ``Kadid-10k: A large-scale artificially
  distorted iqa database,'' in \emph{2019 Eleventh International Conference on
  Quality of Multimedia Experience (QoMEX)}.\hskip 1em plus 0.5em minus
  0.4em\relax IEEE, 2019, pp. 1--3.

\bibitem{jayaraman2012objective}
D.~Jayaraman, A.~Mittal, A.~K. Moorthy, and A.~C. Bovik, ``Objective quality
  assessment of multiply distorted images,'' in \emph{2012 Conference record of
  the forty sixth asilomar conference on signals, systems and computers
  (ASILOMAR)}.\hskip 1em plus 0.5em minus 0.4em\relax IEEE, 2012, pp.
  1693--1697.

\bibitem{xue2013learning}
W.~Xue, L.~Zhang, and X.~Mou, ``Learning without human scores for blind image
  quality assessment,'' in \emph{Proceedings of the IEEE Conference on Computer
  Vision and Pattern Recognition}, 2013, pp. 995--1002.

\bibitem{mittal2012making}
A.~Mittal, R.~Soundararajan, and A.~C. Bovik, ``Making a “completely blind”
  image quality analyzer,'' \emph{IEEE Signal processing letters}, vol.~20,
  no.~3, pp. 209--212, 2012.

\bibitem{zhang2015feature}
L.~Zhang, L.~Zhang, and A.~C. Bovik, ``A feature-enriched completely blind
  image quality evaluator,'' \emph{IEEE Transactions on Image Processing},
  vol.~24, no.~8, pp. 2579--2591, 2015.

\bibitem{li2018has}
D.~Li, T.~Jiang, W.~Lin, and M.~Jiang, ``Which has better visual quality: The
  clear blue sky or a blurry animal?'' \emph{IEEE Transactions on Multimedia},
  vol.~21, no.~5, pp. 1221--1234, 2018.

\bibitem{zhang2021fine}
X.~Zhang, W.~Lin, and Q.~Huang, ``Fine-grained image quality assessment: A
  revisit and further thinking,'' \emph{IEEE Transactions on Circuits and
  Systems for Video Technology}, 2021.

\bibitem{sheikh2006statistical}
H.~R. Sheikh, M.~F. Sabir, and A.~C. Bovik, ``A statistical evaluation of
  recent full reference image quality assessment algorithms,'' \emph{IEEE
  Transactions on image processing}, vol.~15, no.~11, pp. 3440--3451, 2006.

\bibitem{sandler2018mobilenetv2}
M.~Sandler, A.~Howard, M.~Zhu, A.~Zhmoginov, and L.-C. Chen, ``Mobilenetv2:
  Inverted residuals and linear bottlenecks,'' in \emph{Proceedings of the IEEE
  conference on computer vision and pattern recognition}, 2018, pp. 4510--4520.

\bibitem{xie2017aggregated}
S.~Xie, R.~Girshick, P.~Doll{\'a}r, Z.~Tu, and K.~He, ``Aggregated residual
  transformations for deep neural networks,'' in \emph{Proceedings of the IEEE
  Conference on Computer Vision and Pattern Recognition}, 2017, pp. 1492--1500.

\bibitem{zhu2020metaiqa}
H.~Zhu, L.~Li, J.~Wu, W.~Dong, and G.~Shi, ``Metaiqa: deep meta-learning for
  no-reference image quality assessment,'' in \emph{Proceedings of the IEEE/CVF
  Conference on Computer Vision and Pattern Recognition}, 2020, pp.
  14\,143--14\,152.

\bibitem{zhang2022continual}
W.~Zhang, D.~Li, C.~Ma, G.~Zhai, X.~Yang, and K.~Ma, ``Continual learning for
  blind image quality assessment,'' \emph{IEEE Transactions on Pattern Analysis
  and Machine Intelligence}, 2022.

\bibitem{zhang2021task}
W.~Zhang, K.~Ma, G.~Zhai, and X.~Yang, ``Task-specific normalization for
  continual learning of blind image quality models,'' \emph{arXiv preprint
  arXiv:2107.13429}, 2021.

\end{thebibliography}

\end{document}